\documentclass[aps,twocolumn, paper, showpacs,floatfix, nobibnotes,longbibliography]{revtex4-2}
\usepackage{amsmath,amssymb,physics}
\usepackage[utf8]{inputenc}
\usepackage[hypertexnames=false]{hyperref}
\usepackage{datetime}
\usepackage{bm}
\usepackage{enumitem}
\usepackage{graphicx}
\usepackage{braket}
\usepackage{esdiff}
\usepackage{MnSymbol}
\usepackage[T1]{fontenc} 
\usepackage[normalem]{ulem}
\usepackage{mathtools}
\usepackage{bbold}
\usepackage{siunitx}
\usepackage{comment}
\usepackage{soul}
\usepackage{pdfcomment}
\usepackage{gensymb}
\usepackage[dvipsnames]{xcolor}
\usepackage{xfrac}
\usepackage{tikz}
\renewcommand{\selectlanguage}[1]{}
\def\kup{\left\vert\uparrow\right\rangle}
\def\kdn{\left\vert\downarrow\right\rangle}
\def\kaux{\left\vert\mathrm{aux}\right\rangle}

\hypersetup{
    colorlinks,
    linkcolor={red!50!black},
    citecolor={blue!50!black},
    urlcolor={blue!80!black}
}

\begin{document}
\makeatletter
\def\@fnsymbol#1{\ensuremath{\ifcase#1\or \dagger\or \ddagger\or *\or
   \mathsection\or \mathparagraph\or \|\or **\or \dagger\dagger
   \or \ddagger\ddagger \else\@ctrerr\fi}}
\makeatother

\title{Dark state transport between unitary Fermi superfluids}
\author{Mohsen Talebi}

\author{Simon Wili}

\author{Jeffrey Mohan}
\altaffiliation[Present address: ]{Welinq SAS, Paris, France}

\author{Philipp Fabritius}
\altaffiliation{Present address: X-Rite Europe GmbH, Regensdorf, Switzerland}

\author{Meng-Zi Huang}
\email[Corresponding author: ]{mhuang@phys.ethz.ch}

\author{Tilman Esslinger}
\affiliation{Institute for Quantum Electronics and Quantum Center, ETH Z\"urich, 8093 Z\"urich, Switzerland}

\date{2024-11-27}

\begin{abstract} 
The formation of dark states is an important concept in quantum sciences, but its compatibility with strong interparticle interactions---for example, in a quantum degenerate gas---is hardly explored. Here, we realize a dark state in one of the spins of a two-component, resonantly interacting Fermi gas  using a $\Lambda$ system within the $D_2$ transitions of $^6$Li at high magnetic field. The dark state is created in a micrometer-sized region within a one-dimensional channel connecting two superfluid reservoirs. The particle transport between the reservoirs is used as a probe.
We observe that atoms are transported in the dark state and the superfluid-assisted fast current is preserved. If the dark state resonant condition is not met, the transport is  suppressed by the spontaneous emission.
We also uncover an asymmetry in the transport timescale across the two-photon resonance, which is absent in the non-interacting regime and diminished at higher temperatures. This work raises questions on the interplay of dark states with interparticle interactions and opens up perspectives for optical manipulation of fermionic pairing.
\end{abstract}

\maketitle

\emph{Introduction.}---Dark superposition states are the result of quantum interference in a three-level system \cite{cohen-tannoudji_dark_2015, alzetta_experimental_1976}, also giving rise to the phenomenon of electromagnetically-induced transparency~\cite{fleischhauer_electromagnetically_2005}. Applications of this concept include quantum memory in atomic media~\cite{lvovsky2009optical, fleischhauer_dark-state_2000,liu_observation_2001, phillips_storage_2001, wang_efficient_2019, cao_efficient_2020}, compact frequency metrology~\cite{ shah_chapter_2010, knappe_mems_2008}, laser cooling of ultracold atoms and molecules~\cite{cohen-tannoudji_dark_2015, arimondo1996v, ni_high_2008}, and optical control of Feshbach interactions~\cite{thalhammer_inducing_2005, jagannathan_optical_2016}, to name a few. So far, the physics exploited in most systems can be understood at the 
single-atom or molecule level~\cite{winkler_atom-molecule_2005, wu_optical_2012}, with few exceptions such as Rydberg dark-state polaritons where many-body interactions lead to optical nonlinearities~\cite{hofmann_sub-poissonian_2013,zeuthen_correlated_2017}. However, the interplay between atoms' internal dark state and their external degrees of freedom in an interacting system in the quantum degenerate regime has been largely unexplored. In particular, in strongly interacting Fermi gases, 
only molecular dark states that involve fermion pairs as a whole have been realized~\cite{semczuk_anomalous_2014,jagannathan_optical_2016} and studied in regimes where their influence on the many-body pairing at unitarity is still an open question~\cite{semczuk_anomalous_2014}.

Here, in a unitary Fermi gas at low temperatures where many-body pairing leads to superfluid behavior, we experimentally realize a dark state in one of the two spins and  probe it with transport between two superfluid reservoirs. 
The dark state is created by optically  driving $\Lambda$-type transitions locally in a quasi-one-dimensional (quasi-1D) channel connecting the two reservoirs. This configuration therefore protects the superfluid reservoirs from the inevitable residual photon scattering by the $\Lambda$ couplings. In this strongly interacting system, despite residual atom losses in the channel, we observe superfluid-induced transport as a clear signature of the presence of dark states in the channel. 
The dark-state transport provides another probe to the non-equilibrium, many-body system. In particular, the transport timescale across the two-photon detuning reveals a surprising asymmetry absent in the non-interacting regime and diminished at higher temperatures. This optically controlled, spin-dependent dark state also enables local manipulation of one of the two interacting spins, making local engineering of interaction and pairing in fermionic superfluid systems possible.

\nocite{ZurnFeshbach2013}

\begin{figure}[t]
    \includegraphics[width=0.4\textwidth]{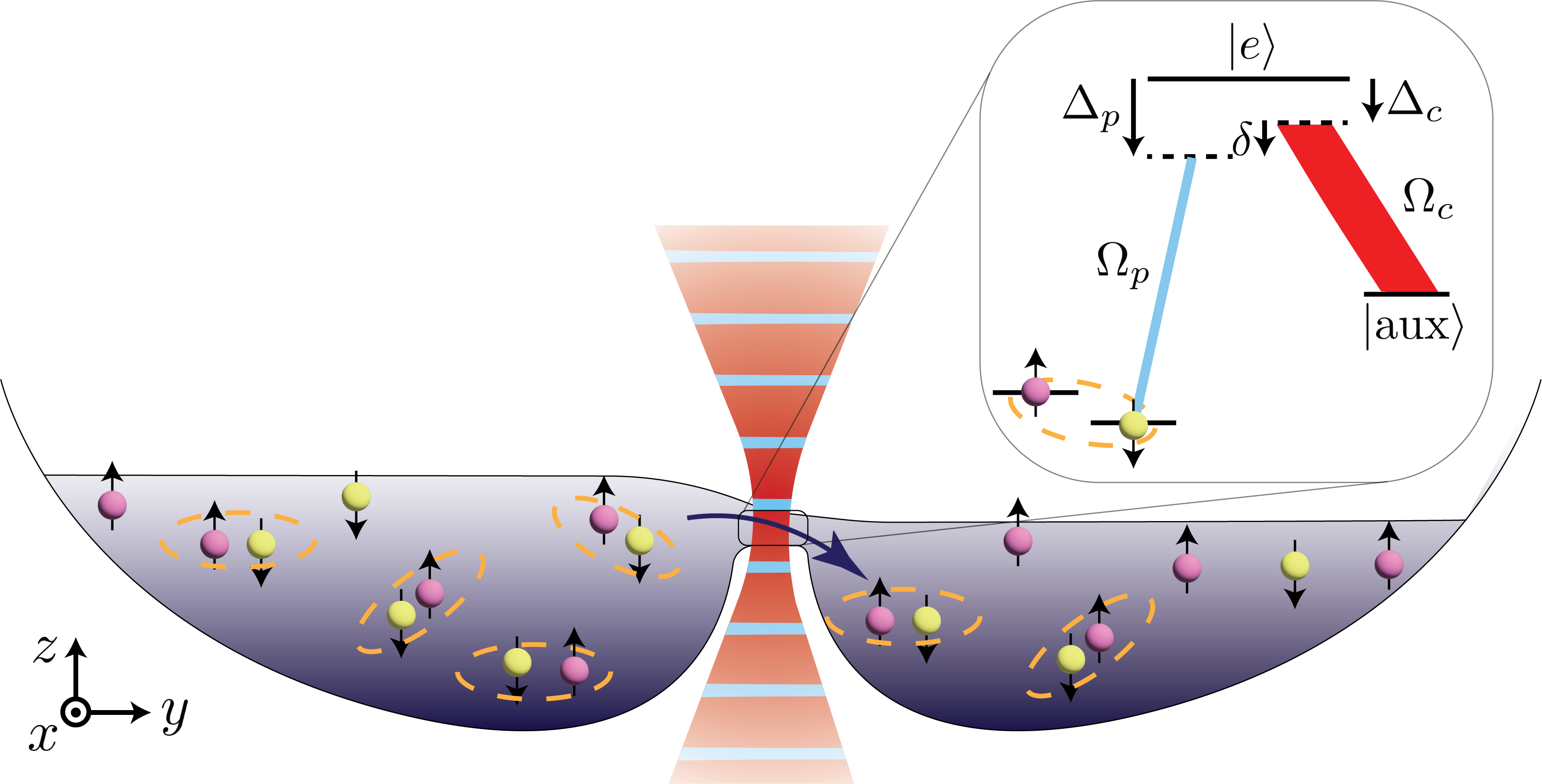}
    \caption{Transport between two unitary Fermi gas reservoirs in which the connecting channel is locally controlled by a dark-state switch acting on spin $\ket{\downarrow}$. Two reservoirs containing a balanced mixture of resonantly interacting atoms in states $\kdn$ and $\kup$ are connected via a quasi-1D transport channel. Inside the 
    channel, two beams with a common spatial mode drive a $\Lambda$ transition (inset), creating a dark state acting like a switch that controls the transport through the channel by the driving fields. The $\Lambda$ scheme couples the atoms in state $\kdn$ to an auxiliary state $\kaux$ via a two-photon process, with the Rabi couplings and detunings defined in the inset. The interactions between $\kaux$ and the two spins are negligible~\cite{supplementary}.}
    \label{fig: figure1}       
\end{figure}

\nocite{mcalexander_radiative_1996}
\nocite{renzoni_coherent_1997}
\nocite{renzoni_population-loss-induced_1998}
\nocite{renzoni_coherent_1999}
\nocite{dalibard_wave-function_1992}
\nocite{molmer_monte_1993}
\nocite{nielsen2010quantum}
\nocite{fleischhauer_electromagnetically_2005}
\nocite{skinner1986pure}
\nocite{streed_continuous_2006}
\nocite{johansson_qutip_2012}
\nocite{huang_superfluid_2023}
\nocite{bartenstein_precise_2005}
\nocite{chin_feshbach_2010}
\nocite{van_abeelen_sympathetic_1997}
\nocite{li_observation_2024}
\nocite{fabritius_irreversible_2024}
\nocite{gehm2003preparation}
\nocite{jones_ultracold_2006}
\nocite{partridge_molecular_2005}

\emph{Experimental setup.}---We prepare a balanced mixture of the first and the third hyperfine ground states of ${}^6\text{Li}$, with pseudo-spin notations $\kdn$ and $\kup$ respectively. The magnetic field is tuned to the Feshbach resonance at $B\approx690~\text{G}$ of the $s$-wave collisions between the two spins \cite{ZurnFeshbach2013}. The gas is confined in a two-reservoir geometry connected via a quantum point contact \cite{krinnerObservationQuantizedConductance2014, husmann_connecting_2015, supplementary}
with transverse confinement frequencies 12.3(2)~kHz and 10.7(1)~kHz along $x$ and $z$, respectively. The transport channel is in the quasi-1D regime since the confinement energy is much higher than the temperature of the cloud $k_BT/h=1.67(3)$~kHz where $k_B$ and $h$ are the Boltzmann and Planck constants. A schematic of the experimental setup is illustrated in Fig.~\ref{fig: figure1}. After preparation of the gas in this geometry, we have a total atom number $N=N_{\downarrow}+N_{\uparrow}= 2.5(2)\times 10^{5}$ with a degeneracy of $T/T_F=0.199(3)$ where $T_F$ is the Fermi temperature of the gas. See Ref.~\cite{supplementary} for more experimental details on the channel's geometry and the experimental cycle.

We implement a local $\Lambda$-scheme inside the 1D channel involving state $\kdn$ and an auxiliary state $|\text{aux}\rangle$ (the fifth hyperfine ground state), see Fig.~\ref{fig: figure1}. The probe and control couplings are facilitated by two co-propagating laser beams of the same spatial mode, which are oriented perpendicularly to the transport direction and focused to a waist of {1.5}~{\micro m} in the channel. In this scheme, a two-photon transition couples the state $\kdn$ to $|\text{aux}\rangle$. The detuning and the effective Rabi frequency averaged over the spatial profile of the probe (control) beam that couples the state $\kdn$ ($|\text{aux}\rangle$) to an excited state $\vert e\rangle$ are denoted with $\Delta_p$ and $\Omega_p$ ($\Delta_c$ and $\Omega_c$), respectively, see the inset of Fig.~\ref{fig: figure1}. The dipole matrix element of the probe transition, $d_p$, is much smaller than that of the control transition, $d_c\approx 18.5 d_p$ ~\cite{gehm2003preparation}, such that $\Omega_p\ll\Omega_c$ for comparable intensities. {As the} two beams are derived from the same laser, their frequency difference is stable on the order of $5~\text{Hz}$. Because of the opposite sign of the magnetic moments of state $\kdn$ and $\vert \text{aux}\rangle$, the magnetic field fluctuation {which is predominantly at low frequencies (such as 50 Hz), }results in an uncertainty of the two-photon detuning $\delta=\Delta_p-\Delta_c$. Because of short transit time through the $\Lambda$ coupling region, each atom only samples the magnetic field noise above 1~kHz (Fourier limit of the transit time). However, the measured signal is sampled from many atoms over seconds and repetitive runs, so is limited by the low-frequency magnetic field noise, leading to a minimally resolved two-photon linewidth $\sim 80$~kHz \cite{supplementary}.

\begin{figure}[t]
    \includegraphics[width=0.5\textwidth]{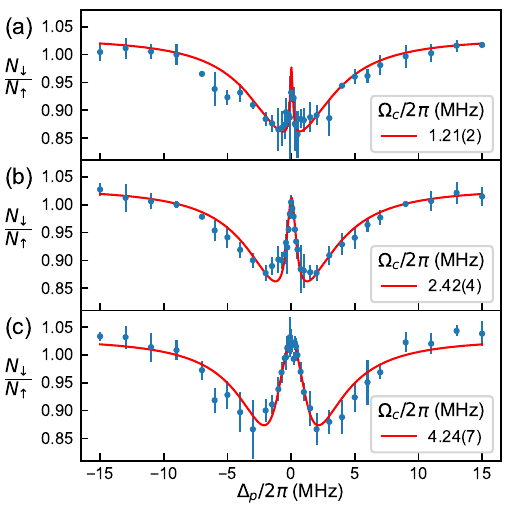}
    \caption{
    Atom loss spectroscopy of the dark-state resonance in the local $\Lambda$ coupling configuration of Fig.~\ref{fig: figure1} with balanced reservoirs. 
    We show $N_\downarrow$ normalized to $N_\uparrow$ as a function of the detuning of the probe, while the control beam is on resonance $\Delta_c=0$. 
    From (a) to (c), increasing $\Omega_c$ shows a broader and more robust dark-state resonance at the center. Error bars represent standard deviation of typically 4--5 repetitions. Solid lines are the results of a simultaneous fit of a semiclassical model to all measured spectra (three out of six shown here)~\cite{supplementary}. 
    In the fit, the common $\Omega_p$ and all $\Omega_c$'s are constrained using the known control-probe power ratios, yielding $\Omega_p=2\pi\times52(1)$~kHz and $\Omega_c$ shown in each plot \cite{supplementary}.
    \label{fig: figure2}}       
\end{figure}

\emph{Dark-state resonance by loss spectroscopy.---} 
We first demonstrate the presence of the dark-state resonance by measuring the atom loss as a function of the probe detuning. For this, we prepare the reservoirs at identical thermodynamic conditions and measure the total atom number in each spin after $2~\text{s}$ of illumination of the $\Lambda$ couplings in the channel. Without the control beam, a resonant probe beam excites $\kdn$ atoms into $\vert e \rangle$ which quickly decay to $\kaux$ (since $d_c\gg d_p$) and leave the system, resulting in loss of $N_\downarrow$. Scanning the detuning $\Delta_p$ of a weak probe reveals the broad loss resonance with a linewidth mainly limited by the decay rate of the excited state, $\Gamma_e\approx 2\pi \times 6$~MHz. With the control beam on and $\Delta_c\approx 0$, a peak in $N_\downarrow$ appears at $\Delta_p=0$ due to the existence of a dark state 
which is decoupled from the laser fields and suppresses the atom loss due to scattering. In Fig.~\ref{fig: figure2} we show the measurements revealing the dark resonance at various choices of $\Omega_c$. Here we plot $N_\downarrow$ normalized to $N_\uparrow$ to eliminate the effect of shot-to-shot atom number fluctuations that are common to both spins. The observed width of the dark-state resonance is  limited by the magnetic field noise, as well as the intensity broadening by $\Omega_p$ and $\Omega_c$. 
By increasing $\Omega_c$ we observe a broader resonance but also a more robust dark state as is expected also in the absence of collisional interactions \cite{fleischhauer_electromagnetically_2005}. Two additional loss channels affect the observed spectra in $N_\downarrow/N_\uparrow$: 
1) loss of $\kup$ due to the strong interaction between the spins when a $\kdn$ atom is {excited and subsequently lost} \cite{huang_superfluid_2023}.  2) pair losses induced by the control beam due to a few off-resonant photoassociation transitions \cite{supplementary}.

In our experiment, the photon recoil energy $E_R/k_B\approx 3.5~\micro$K is much larger than the trap depth $U_0/k_B\approx 1.2~\micro$K, hence an atom is lost with near-unity probability after a single scattering event. 
Therefore, an open-system treatment is needed to find out how the atoms survive by evolving into the dark state without spontaneous emission, in contrast to the commonly used closed-system treatment~\cite{fleischhauer_electromagnetically_2005}. Our model assumes particle loss after each spontaneous emission event~\cite{supplementary} using the Monte Carlo wave-function approach~\cite{dalibard_wave-function_1992, molmer_monte_1993}. The loss mechanism damps the atoms into the dark state, similar to the process of coherent population trapping~\cite{shore1991theory}.
We access the density matrix of a surviving atom by \emph{conditioning} the evolution on no spontaneous emission. 
Although these atoms do not undergo spontaneous emission, they inherit the projective nature of the emission via the conditioning, similar to the phenomenon of the quantum Zeno effect \cite{streed_continuous_2006}. We use the steady-state solution of the excited state population from the conditioned evolution to obtain the particle loss rate in the $\Lambda$ region.
A phenomenological model of atoms passing through the $\Lambda$ region with this loss rate allows us to fit the measured loss spectra (solid lines in Fig.~\ref{fig: figure2}). These fits are performed simultaneously for all six measured spectra varying $\Omega_c$ \cite{supplementary} with three free parameters, including calibration factors of the effective Rabi frequencies and laser frequencies.
Three of these spectra are plotted in Fig.~\ref{fig: figure2}. The fit result $\Omega_p=2\pi\times 52(1)$~kHz is in good agreement with the measured probe Rabi coupling averaged over the Gaussian waist, $2\pi\times58(2)$~kHz~\cite{supplementary}. Control $\Omega_c$'s are constrained to $\Omega_p$ in the fit using the measured power ratios and the known ratio of the dipole matrix elements $d_c/d_p$. In these {spectroscopic} measurements, {there is no signature of} the strong interaction between $\kdn$ and $\kup$ {within our} measurement precision. 

\begin{figure}[t]
    \includegraphics[width=0.5\textwidth]{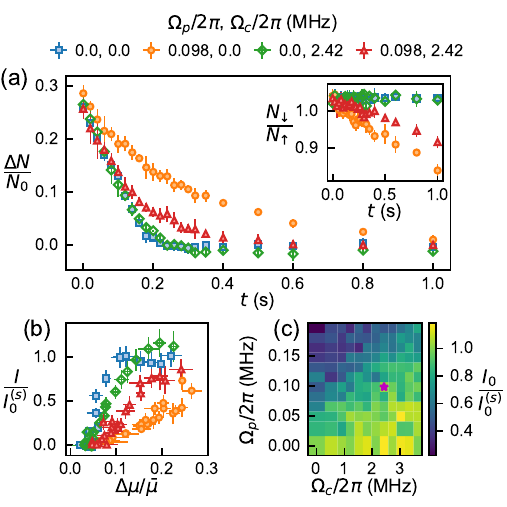}
    \caption{ Dark state recovers superfluid characters of the transport. (a) Time evolution of the particle imbalance 
    for probe and control beam of different strengths. The numerically extracted current-bias relation for the same data are shown in (b). Without any $\Lambda$ coupling (blue squares), the imbalance shows a fast non-exponential decay, corresponding to a non-Ohmic current-bias relation in (b). With only a resonant probe beam of $\Omega_p=2\pi\times 98~\text{kHz}$ (orange circles), the induced dissipation in the channel leads to a much slower transport as well as a nearly Ohmic current-bias relation. Keeping the same probe and turning on a resonant control $\Omega_c=2\pi\times2.42~\text{MHz}$ turns $\kdn$ atoms into the dark state (red triangles). The fast current, as well as its nonlinear characteristics, are recovered to a large extent. The control beam alone (green diamonds) has negligible effect on the transport, only slightly reducing the nonlinearity, visible in (b). (c) Initial current obtained from a linear fit to $\Delta N(t)$ for transport time $t\leq0.08~\text{s}$ as a function of probe and control Rabi frequencies. All the currents in (b,c) are normalized to that for $\Omega_p=\Omega_c=0$. The magenta star marks the condition for data in (a,b). 
    \label{fig: figure3} }      
\end{figure}

\emph{Probing the dark state with superfluid transport.---}The observed dark state is a coherent superposition of $\kdn$ and $\kaux$: $\psi_D \propto \Omega_c\kdn - \Omega_p\kaux$, which has a majority population in $\kdn$ given $\Omega_c\gg\Omega_p$ in our setting. We now address the question of whether the pairing between $\kdn$ and $\kup$ in the unitary superfluid is preserved under the local conversion into this dark state.
The nature of the gas can be probed by the particle transport between the reservoirs, where superfluidity results in a distinctive non-Ohmic current beyond linear response \cite{husmann_connecting_2015, fabritius_irreversible_2024}. We measure the current {by first preparing} an initial atom number imbalance $\Delta N$ between the two reservoirs with $\Delta N/N_0=0.27(1)$ ($N_0$ is the initial total atom number in the cloud). Subsequently, we observe its evolution in time. The relaxation of $\Delta N/N_0$ for various settings of $\Omega_p$ and $\Omega_c$ is presented in Fig.~\ref{fig: figure3}(a). In the absence of the local $\Lambda$ couplings, $\Omega_p=\Omega_c=0$, we observe a fast non-exponential relaxation of the imbalance between the two reservoirs. The corresponding apparent current, $I=-(1/2)\dv*{\Delta N}{t}$, is non-Ohmic, as evidenced by the nonlinear current-bias relation [blue squares in Fig.~\ref{fig: figure3}(b)], indicating the superfluid character of the transport process 
\cite{husmann_connecting_2015}. When we turn on a resonant probe with $\Omega_p = 2\pi\times 98$~kHz without the control beam, the particle current is reduced and shows nearly Ohmic behavior [orange circles in Figs.~\ref{fig: figure3}(a),(b)]. This is due to the presence of atom losses in the channel as observed previously  \cite{huang_superfluid_2023}.

Next, we create the dark-state condition in the channel by adding a resonant strong control beam with Rabi coupling $\Omega_c = 2\pi\times2.42 \ \text{MHz}$. Now the transport is fast again, with the initial current and the nonlinearity being largely recovered [red triangles in Figs.~\ref{fig: figure3}(a),(b)]. This observation stands as an evidence for the transport of $\kdn$ in a dark state while the process largely preserves pairing and the superfluid character. The reduction of the spin-dependent loss is also visible from the time evolution of the $N_\downarrow/N_\uparrow$ [inset of Fig.~\ref{fig: figure3}(a)]. In the final ``control'' experiment, we only keep on the control beam with the same $\Omega_c = 2\pi\times2.42 \ \text{MHz}$ while turning off the probe. Despite pair losses induced by the control beam \cite{supplementary}, the $\Delta N$ decay,  hence the {apparent} current, is almost indistinguishable from the case without any $\Lambda$ coupling [green diamonds in Fig.~\ref{fig: figure3}(a,b)]. Therefore, the effect of the pair losses {on the apparent current} is negligible. 

To further investigate the robustness of the dark-state transport, we sample a wide range of both the probe and control Rabi frequencies. We determine the initial particle current $I_0$ with a linear fit to $\Delta N(t)$ within short transport time $t\leq0.08$~s. In Fig.~\ref{fig: figure3}(c), we plot $I_0$ for combinations of $\Omega_p$ and $\Omega_c$, normalized to $I_0^s$ that corresponds to the case of $\Omega_p=\Omega_c=0$. For a given choice of $\Omega_p$, as we increase $\Omega_c$, we observe a trend of increasing initial current, reaching $I_0^s$. As we go to higher values of $\Omega_p$,  higher $\Omega_c$ is required to maintain a fast transport. At the same time, we observe an increased pair loss by the control beam when its power is increased, which eventually suppresses the current. The configuration for Figs.~\ref{fig: figure3}(a),(b) is chosen in the intermediate regime [denoted by the star in Fig.~\ref{fig: figure3}(c)].
Despite that the amplitude of $\ket{\mathrm{aux}}$ in the dark state is small ($\Omega_p/\Omega_c$) in the explored regimes, the pairing between $\ket{\downarrow}$ and $\ket{\uparrow}$ could in principle be affected by such perturbation in analogy to broad Feshbach resonances originating from coupling to a tiny molecular fraction~\cite{partridge_molecular_2005} and the optically controlled Feshbach resonances~\cite{wu_optical_2012} utilizing small admixture of closed-channel molecular states~\cite{arunkumar_designer_2019}.

\begin{figure}[t]
    \includegraphics[width=0.5\textwidth]{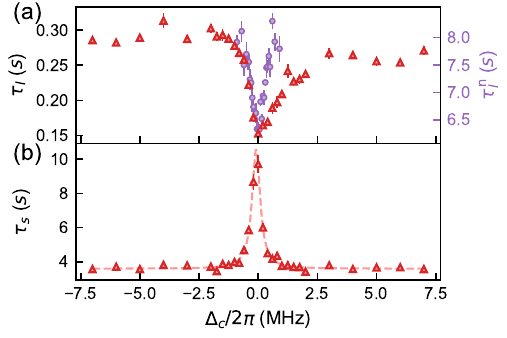}
    \caption{Observation of asymmetric transport timescale versus two-photon detuning in the unitary gas. (a) The transport timescale as a function of the two-photon detuning (varying $\Delta_c$ while $\Delta_p=0$) in the unitary regime (red triangles, left axis), with the same $\Omega_p$ and $\Omega_c$ as in Figs.~3(a),(b), red triangles. Each point is the time constant $\tau_I$ of an exponential decay, $\mathcal{A}e^{-t/\tau_I}$, fitted to the evolution of the imbalance ${\Delta N}/{N_0}$, with $\mathcal{A}$ fixed to the average of all data at $t=0$. 
    We observe that the characteristic transport time of the unitary gas responds asymmetrically to the sign of two-photon detuning. In contrast, a separate measurement in the non-interacting regime (purple circles, right axis) taken under similar $\Lambda$ couplings [$\Omega_p=2\pi\times 112(1)$~kHz, $\Omega_c= 2\pi\times 2.62(3)$~MHz] does not show such asymmetry. Time constants $\tau_I^\text{n}$ for the non-interacting regime are obtained in the same way as $\tau_I$ but fitting $\Delta N_\downarrow/N_\uparrow$ where $N_\uparrow$ simply serves as a precise estimate of $N_\downarrow(0)$. (b) The corresponding loss rate versus two-photon detuning in the unitary regime shows no asymmetry [same data as the red triangles in (a)]. The time constants $\tau_s$ are obtained from exponential fits $\bar{s_0}e^{-t/\tau_s}$ to the spin ratio $s=N_\downarrow/N_\uparrow$, with  $\bar{s_0}$ fixed to the average of all data at $t=0$. The dashed line is a Lorentzian fit as a guide to the eye. Error bars represent standard errors from the fits.}
    \label{fig: figure4}       
\end{figure}

\emph{Asymmetric transport versus two-photon detuning.---} We further investigate the behavior of the transport in terms of the two-photon detuning close to the dark-state resonance. We perform the same transport measurement as above with a detuned control frequency $\Delta_c\in 2\pi\times[-7,7]$~MHz while keeping the probe frequency on the single photon resonance ($\Delta_p=0$). The effective Rabi couplings $\Omega_p$ and $\Omega_c$ are chosen the same as previously in Fig.~\ref{fig: figure3}(a), red triangles. 
We characterize the transport by the timescale $\tau_I$ obtained from an exponential fit  to $\Delta N/N_0$.  The results versus the control detuning are shown in Fig.~\ref{fig: figure4}(a), red triangles. Here, we observe a clearly visible asymmetry of the characteristic transport times across the resonance: the particle imbalance relaxes faster for $\Delta_c>0$ than $\Delta_c<0$ for a given $\abs{\Delta_c}$ near resonance. 
However, we do not observe such a behavior in the atom loss rate, which is symmetric with respect to $\Delta_c$ as characterized by the decay of the spin ratio $s(t)=N_\downarrow(t)/N_\uparrow(t)$ (as used in Fig.~\ref{fig: figure2}). 
This is visible from exponential-decay fits to $s(t)$, whose decay times $\tau_s$ are plotted versus $\Delta_c$ in Fig.~\ref{fig: figure4}(b). At the single-atom level, a possible origin of an asymmetry is the dispersive asymmetry in the local conservative potential of the $\Lambda$-coupling away from the two-photon resonance. The observed asymmetry suggests a repulsive (attractive) potential with a negative (positive) $\Delta_c$, resulting in a slower (faster)  transport. 
However, within our conditioned evolution description, atoms experience a negligible net light shift in the steady state \cite{supplementary}. Furthermore, the simulated transient evolution before reaching the steady state shows that the light shift is maximally $k_B \times 4~\mathrm{nK}$ for the explored range of $\Delta_c$, negligible compared to any relevant energy scales such as the temperature ($\sim80~\text{nK}$) or Fermi energy ($\sim k_B \times 400~\text{nK} $).
Our theory of three-level single-particle light-matter interaction does not seem to explain the observed asymmetry, yet the theory is consistent with  a separate measurement in the non-interacting regime where we do not observe such an asymmetry [purple circles in Fig.~\ref{fig: figure4}(a)]. 
We can further exclude dipole potentials due to other off-resonant transitions~\cite{supplementary}.
In principle, the asymmetry could still originate from single-particle behavior but is too small to measure in the non-interacting regime while the superfluid-enhanced current amplifies the subtle effect. One possible mechanism could be non-adiabatic conservative potentials \cite{dum1996gauge} during the transient atomic evolution before reaching the steady state. 

On the other hand, the asymmetry could also have a many-body origin. We also observe that the asymmetry diminishes when performing the experiment at unitarity but higher temperatures ($T/T_F\approx 0.77$ in the reservoirs, data similar to the non-interacting case but transport is still much faster; see Ref.~\cite{supplementary}), further suggesting that the asymmetry is related to pairing between $\kdn$ and $\kup$.
One possible mechanism is similar to the asymmetric rf spectrum of the pairing gap~\cite{Chin_pairing_2004, chin_radio-frequency_2005, schunck_determination_2008}.
Here, the $\Lambda$ scheme couples the resonantly interacting $\kdn-\kup$ pairs to the free $\vert\text{aux}\rangle-\kup$ pairs. The pairing between $\kdn$ and $\kup$ leads to nonzero two-photon coupling to a broad continuum of relative momentum states of the free pairs for $\Delta_c>0$, effectively broadening the dark-state resonance on the $\Delta_c>0$ side. This picture is consistent with the observed faster transport at $\Delta_c>0$, but also predicts an asymmetry in the atom loss. It is possible that the latter exists below our experimental resolution [Fig.~\ref{fig: figure4}(b)], highlighting that transport in the unitary gas is a sensitive probe of many-body pairing.
A detailed study of these scenarios goes beyond the scope of this work as it requires new approaches or extending the theoretical studies on interacting fermions in spin-dependent $\Lambda$ schemes \cite{dao_measuring_2007, jiang_detection_2009, zhou_phase_2012, jen_theory_2013} to transport settings.

\emph{Conclusion.---}
In this work, we reported on an experimental realization of a spin-selective dark state inside a 1D channel connecting two fermionic unitary superfluid reservoirs. We observed a revival of the non-Ohmic current-bias characteristic as a result of the dark-state transport through the channel. Furthermore, we observed an asymmetric dependence of the characteristic transport time on the two-photon detuning. This asymmetry is not observed in the non-interacting regime, suggesting that a full explanation might require taking into account many-body effects. Our work paves the way for studying dark states in the strongly interacting regime and their interplay with superfluid transport phenomena. 
This spin-selective $\Lambda$ scheme also enables various local spin manipulations in strongly interacting systems, such as \emph{spatial} stimulated Raman adiabatic passage with spatially separated $\Lambda$ coupling beams to arbitrarily control the dark-state composition~\cite{Vitanov2017,Bergmann_2019}, and spin-dependent subwavelength potentials \cite{wang_dark_2018, lacki_nanoscale_2016}.

\emph{Acknowledgments}---
We thank Alexander Frank for technical support. 
We are grateful for discussions with Alex G\'omez Salvador and Eugene Demler. We also thank Ran Qi for sharing his calculation of the scattering lengths. We acknowledge the Swiss National Science Foundation (Grants No.~212168, UeM019-5.1, and TMAG-2~209376) and European Research Council advanced grant TransQ (Grant No.~742579) for funding. 
\bibliography{paper}

\clearpage

\section*{Supplemental material}
\setcounter{figure}{0}
\renewcommand{\thefigure}{S\arabic{figure}}

\section{Formalism of three-level system with particle loss}
\label{sec: Conditional Lambda system}

In this section, we formulate the system's density matrix under $\Lambda$ coupling while taking into account atom loss due to the recoil of photon scattering. This is done by considering two subspaces of external states, one surviving and one lost, and the jump processes between them. We arrive at the nonlinear evolution equation of the surviving atoms. On the one hand, we use a numerical method to extract the state of the surviving atoms for an arbitrary time. On the other hand, we analytically solve for the steady state of this evolution, which is later used in fitting the measured spectra.

\subsection{Derivation of the state evolution of a surviving atom}

The $\Lambda$ system we use couples the first ($\kdn=\vert 1\rangle$) and the fifth ($\vert\text{aux}\rangle=\vert 5\rangle$) hyperfine ground states to a common excited state $\vert e\rangle=\vert 2 P_{3/2}, m_J=3/2, m_I=0\rangle$ using two  $\sigma_+$ polarized beams with a common spatial mode in the $D_2$ transition manifold. The total {decay rate of the excited state} is {$\Gamma_e=2\pi\times5.87\ \text{MHz}$}  \cite{mcalexander_radiative_1996} with dominant decay rates to $\ket{1}$ and $\ket{5}$ of {$\Gamma_{e1}\approx 2.9\times10^{-3} \Gamma_e$} and {$\Gamma_{e5}\approx0.997 \Gamma_e$}, respectively. The total spontaneous decay rate to other states is {$\sim 8\times 10^{-7} \Gamma_e$} {so} that we can neglect atomic transitions outside {the} $\vert1\rangle-\vert e\rangle-\vert 5\rangle$ system.

In our experimental condition, a comparison of the recoil energy {$E_r=\hbar^2 k^2/(2m)\approx k_B\times 3.5 \ \micro \text{K}$} and the trap depth {$U_0\approx k_B \times 1.2 \ \micro \text{K}$} shows that an atom is lost with a high chance after a single spontaneous emission event.  Therefore, the evolution of the surviving atoms in the cloud needs to be modeled by considering a loss channel outside of the $\Lambda$ system. As discussed in \cite{renzoni_coherent_1997, renzoni_population-loss-induced_1998, renzoni_coherent_1999}, the behavior of the coherent population trapping in the dark state can be influenced dramatically by such a loss process. 
To model such a loss mechanism, we adapt the physical picture of gedanken projective measurements in the Monte Carlo wavefunction (MCWF) approach \cite{dalibard_wave-function_1992, molmer_monte_1993} and simultaneously model the internal and external degrees of freedom of a single atom. As we detail below, the major difference of our derivation compared to the Refs.~\cite{dalibard_wave-function_1992, molmer_monte_1993} lies in \emph{conditioning} the survival of an atom on the absence of spontaneous emission. We will show that this assumption results in a nonlinear master equation of the survived atoms.

Assume that the atom is in the pure product state $\vert \psi\rangle=\vert \phi_{\text{int}} \rangle \otimes \vert \phi_{\text{ext}} \rangle$ where $\vert \phi_{\text{int}} \rangle$ and $\vert \phi_{\text{ext}} \rangle$ denote the internal atomic and external motional states. We classify the external state in two categories:
\begin{align}
    & S_{\text{in}}= \text{span}\left\{\vert \bm{p} \rangle \colon  \abs{\bm{p}}<p_c  \right\},  \\
    & S_{\text{out}}= \text{span}\left\{\vert \bm{p} \rangle \colon  \abs{\bm{p}}>p_c  \right\},
\end{align}
where $S_{\text{in}}$ is the set of states that survive in the $\Lambda$ region while $S_{\text{out}}$ is the set of states that are eventually lost from the trap. Here, we assume that an atom is lost from the trap if it has a momentum higher than the critical momentum $p_c$ determined by the trap depth $p_c \sim \sqrt{2mU_0}$. Since the recoil momentum $\hbar k$ is much larger than $p_c$, we assume that a spontaneous emission event {transfers an atom} initially in $ S_{\text{in}}$ to $S_{\text{out}}$. 

Assume that the state of the atom at time $t$ is $\vert \Psi(t) \rangle= \vert \phi_{\text{int}}(t) \rangle \otimes \vert \phi_{\text{ext}}^{(\text{in})}(t) \rangle$ in which $\vert \phi_{\text{int}}(t) \rangle =\alpha_1 \vert 1 \rangle + \alpha_5 \vert 5 \rangle + \alpha_e \vert e \rangle$ and $\vert \phi_{\text{ext}}^{(\text{in})}(t) \rangle \in S_{\text{in}}$. 
The Hermitian part of the evolution {in the $\Lambda$ region is governed by the Hamiltonian $\hat{H}=\hat{H}_{\Lambda} \otimes \hat{\Pi}_{\text{in}} + \hat{H}_\Lambda ' \otimes \hat{\Pi}_{\text{out}}$} in which $\hat{\Pi}_{\text{in}}$ ($\hat{\Pi}_{\text{out}}$) is the projection operator into $S_{\text{in}}$ ($S_{\text{out}}$). While $\hat{H}_\Lambda '$ might differ from $\hat{H}_\Lambda$, only the latter, the light-matter Hamiltonian of the surviving atoms, is relevant in the following.  The Hamiltonian $\hat{H}_\Lambda$ in the rotating frame with rotating wave-approximation reads 
    \begin{align}
    &\hat{H}_\Lambda=-\frac{\hbar}{2} 
    \begin{pmatrix}
    &0 &0 &\Omega_p \\
    &0 &-2(\Delta_p- \Delta_c)&\Omega_c \\
    &\Omega_p &\Omega_c &-2\Delta_p
    \end{pmatrix}, \label{eq: Hamiltonian in}
    \end{align}
where $\Delta_p=\omega_{1e}-\omega_p$ and $\Delta_c=\omega_{5e}-\omega_c$ denote the probe and control detuning respectively, and $\omega_p$ and $\omega_c$ are the laser frequencies and $\omega_{1e}$ and $\omega_{5e}$ are the  frequencies of the corresponding atomic transitions. Furthermore in \eqref{eq: Hamiltonian in}, the Rabi couplings of the probe and control for the survived atoms in $S_{\text{in}}$ are denoted by $\Omega_p$ and $\Omega_c$ respectively.

\subsubsection*{Jump processes and probabilities}
During a differential time evolution $dt$, we consider the following Markovian jump processes in the framework of MCWF:
\begin{enumerate}
    \item Spontaneous emission from $\vert e \rangle$ to $\vert 5 \rangle$ or $\ket{1}$ with jump operators
    \begin{align}
        \hat{\mathcal{S}}^{\text{spon}}_{e5}=\sqrt{\Gamma_{e5}}  \vert 5 \rangle \langle e \vert \otimes \vert \phi_\text{ext}^{(\text{out})}(t) \rangle \langle \phi_\text{ext}^{(\text{in})}(t) \vert, \label{eq: spont e-5 in-out jump}\\
        \hat{\mathcal{S}}^{\text{spon}}_{e1}=\sqrt{\Gamma_{e1}}  \vert 1 \rangle \langle e \vert \otimes \vert \phi_\text{ext}^{(\text{out})}(t) \rangle \langle \phi_\text{ext}^{(\text{in})}(t) \vert, \label{eq: spont e-1 in-out jump}
    \end{align}
    which simultaneously change the external state of the atom to $\vert \phi_\text{ext}^{(\text{out})}(t) \rangle \in S_{\text{out}}$ encoding the particle loss due to the spontaneous emission. These jump processes occur with probabilities $dp_{e5}^{\text{spon}}=dt\langle \Psi\vert (\hat{\mathcal{S}}^{\text{spon}}_{e5})^{\dagger} \hat{\mathcal{S}}^{\text{spon}}_{e5} \vert \Psi \rangle =\Gamma_{e5} \vert \alpha_e \vert^2 dt$ {and $dp_{e1}^{\text{spon}}=dt\langle \Psi\vert (\hat{\mathcal{S}}^{\text{spon}}_{e1})^{\dagger} \hat{\mathcal{S}}^{\text{spon}}_{e1} \vert \Psi \rangle =\Gamma_{e1} \vert \alpha_e \vert^2 dt$, respectively,} during a differential time evolution $dt$.

    \item Dephasing due to fluctuations of the two-photon detuning $\delta=\Delta_p-\Delta_c$. In our experiment, the major source of this non-Hermitian process is the magnetic field noise.  From the Hamiltonian \eqref{eq: Hamiltonian in}, the Markovian fluctuations in the two-photon detuning will result in random phase kicks \cite{nielsen2010quantum} modelled by a jump operator of the form \cite{fleischhauer_electromagnetically_2005, molmer_monte_1993}
    \begin{equation}
        \hat{\mathcal{S}}^{\text{deph}}_{5}=\sqrt{\gamma_{5}}  \vert 5 \rangle \langle 5 \vert \otimes \hat{I}^{\text{ext}}, \label{eq: dephasing 5 jump}
    \end{equation}
    which projects the internal state to $\vert 5 \rangle$ while not influencing the external state as a result of the identity operator $\hat{I}^{\text{ext}}$. We assume that the two-photon detuning is of the form $\delta=\tilde{\delta}+\eta(t)$ in which $\tilde{\delta}$ is the average value and $\eta(t)$ is sampled from a white Gaussian noise with zero mean. With this assumption, we can show that the system undergoes the jump process \eqref{eq: dephasing 5 jump} if the autocorrelation function of the white noise is set to $\langle \eta(t) \eta(t^\prime)\rangle=\gamma_5 \delta_D(t-t^\prime)$ with $\delta_D$ denoting the Dirac's $\delta$-function \cite{skinner1986pure}. The probability of an occurrence of the jump during the differential time $dt$ is $dp_{5}^{\text{deph}}=dt\langle \Psi\vert (\hat{\mathcal{S}}^{\text{deph}}_{5})^{\dagger} \hat{\mathcal{S}}^{\text{deph}}_{5} \vert \Psi \rangle =\gamma_{5}\abs{\alpha_5}^2 dt$. We mention that this dephasing process results in a reduction of the peak value of the EIT spectrum for small enough $\Omega_p$ and $\Omega_c$ [see Fig.~2(a)]. Furthermore, in cases that this dephasing process is the limiting factor ($\Omega_p\ll \Omega_c\sim\gamma_5$),  $\gamma_5$ describes the full width at half maximum (FWHM) of the EIT resonance \cite{skinner1986pure}. We determined an upper bound $\gamma_5<2\pi\times 80 \ \text{kHz}$ by measuring the linewidth of this resonance at small Rabi couplings in a non-interacting gas. In {the theoretical model}, we fix $\gamma_5$ to this measured upper bound.  
    
    \item Dephasing due to the fluctuations of the single-photon probe detuning $\Delta_p$. In our experiment, the major source of this non-Hermitian process is the uncertainty in the absolute {laser frequency, which} do not affect $\delta$ because both the probe and control beams are derived from the same laser source. From the Hamiltonian \eqref{eq: Hamiltonian in}, the Markovian fluctuations in the single-photon detuning will result in a jump operator of the form \cite{nielsen2010quantum, fleischhauer_electromagnetically_2005, molmer_monte_1993}
    \begin{equation}
        \hat{\mathcal{S}}^{\text{deph}}_{e}=\sqrt{\gamma_{e}}  \vert e \rangle \langle e \vert \otimes \hat{I}^{\text{ext}}, \label{eq: dephasing e jump}
    \end{equation}
    which projects the internal state to $\vert e \rangle$ while not influencing the external state. The probability of an occurrence of this jump during the differential time $dt$ is $dp_{e}^{\text{deph}}=dt\langle \Psi\vert (\hat{\mathcal{S}}^{\text{deph}}_{e})^{\dagger} \hat{\mathcal{S}}^{\text{deph}}_{e} \vert \Psi \rangle =\gamma_{e}\abs{\alpha_e}^2 dt$. We fix this dephasing {rate in the model} to the typical standard deviation of the laser  frequency {(locked with a wavelength meter)} $\gamma_e\sim 2\pi\times 1.5\ \text{MHz}$.
\end{enumerate}

\subsubsection*{Conditioned state evolution after jump processes}
Conditioned on not losing the atoms from the system (no spontaneous emission events \eqref{eq: spont e-5 in-out jump} and \eqref{eq: spont e-1 in-out jump}), the atom will be in one of the following states after the time evolution $dt$:
\begin{enumerate}
    \item \textbf{Evolution without any jump:} With probability $\mathcal{P}_1=(1-dp_{5}^{\text{deph}})(1-dp_{e}^{\text{deph}})$, the atom will not undergo any {dephasing} \eqref{eq: dephasing 5 jump} and \eqref{eq: dephasing e jump}. Under this condition, the state of the atom can be calculated with the non-Hermitian evolution
    \begin{align}
        \vert\Psi_1(t+&dt) \rangle= \nonumber\\
        &\zeta\left[1-\frac{i}{\hbar}  dt (\hat{H}-\frac{i \hbar}{2} \sum_{j} \hat{\mathcal{S}}_j^\dagger \hat{\mathcal{S}}_j ) \right] \vert \Psi(t) \rangle, \label{eq: MCWF evolution 1}
    \end{align}
    where $S_j$ are the jump operators \eqref{eq: spont e-5 in-out jump}, \eqref{eq: spont e-1 in-out jump}, \eqref{eq: dephasing 5 jump}, and \eqref{eq: dephasing e jump} and $\zeta$ is a normalization constant. Note that although we condition the problem on no spontaneous emission events \eqref{eq: spont e-5 in-out jump} and \eqref{eq: spont e-1 in-out jump}, the projective nature of the absence of these events appears in the non-Hermitian contributions in the evolution \eqref{eq: MCWF evolution 1}. In fact, since our experimental parameters are in the limit of $\Gamma_e\gg \Omega_p$, this projective nature results in a reduction of the particle loss due to the continuous quantum Zeno effect \cite{streed_continuous_2006}. {One can show that}
    \begin{equation}
        \zeta=\prod_j \dfrac{1}{\sqrt{1-dp_j}}, \label{eq: normalization MCWF tot}
    \end{equation}
    where $dp_j$ denotes the probability of the occurrence of each jump process. By inserting all the jump operators in equation \eqref{eq: MCWF evolution 1}, the state of the atom is of the form $\vert \Psi_1(t+dt) \rangle= \vert \phi_1(t+dt) \rangle \otimes \vert \phi_\text{ext}^{(\text{in})}(t+dt)\rangle$ where the atom survives in the system ($\vert \phi_\text{ext}^{(\text{in})}(t+dt) \rangle \in S_{\text{in}}$). The resulting form of the internal state is:
    \begin{align}
        \vert \phi_1(t+dt)\rangle = 
        &\zeta \bigg[\vert \phi(t)\rangle - \frac{i}{\hbar}\hat{H}_{\text{in}} \vert \phi(t)\rangle dt \nonumber \\
        &- \frac{\gamma_5}{2} \alpha_5 \vert 5 \rangle dt - \frac{\Gamma_e+\gamma_e}{2} \alpha_e \vert e\rangle dt \bigg],
    \end{align}
    with $\Gamma_e=\Gamma_{e1}+\Gamma_{e5}$, the total linewidth of the excited state.
    \item \textbf{{A jump due to two-photon dephasing}:} With probability $\mathcal{P}_2=dp_5^{\text{deph}}$, the atom will undergo the dephasing jump \eqref{eq: dephasing 5 jump}, after which it will collapse to the state
    \begin{equation}
        \vert \Psi_2(t+dt) \rangle = \underbrace{\vert 5\rangle}_{\vert \phi_2 \rangle} \otimes \vert \phi_{\text{ext}}^{(\text{in})}(t+dt)\rangle. \label{eq: MCWF evolution 2}
    \end{equation}
    \item \textbf{{A jump due to single-photon dephasing}:} With probability $\mathcal{P}_3=dp_e^{\text{deph}}$, the atom will undergo the dephasing jump \eqref{eq: dephasing e jump} and collapses to the state
    \begin{equation}
        \vert \Psi_3(t+dt) \rangle = \underbrace{\vert e\rangle}_{\vert \phi_3 \rangle} \otimes \vert \phi_{\text{ext}}^{(\text{in})}(t+dt)\rangle. \label{eq: MCWF evolution 3}
    \end{equation}
    \item \textbf{Simultaneous jumps of both single and two-photon dephasing:} Both of the jump events \eqref{eq: dephasing 5 jump} and \eqref{eq: dephasing e jump} can happen simultaneously with the probability $\mathcal{P}_4=dp_5^{\text{deph}}dp_e^{\text{deph}}$. However, since both $dp_5^{\text{deph}}$ and $dp_e^{\text{deph}}$ are of order $dt$, the probability $\mathcal{P}_4$ will be of order $dt^2$ and {is} neglected.
\end{enumerate}

\subsubsection*{Master equation of the surviving atom}
After the time evolution $dt$, the state of the system conditioned on survival inside the trap will be the statistical mixture of the internal part of states \eqref{eq: MCWF evolution 1}, \eqref{eq: MCWF evolution 2}, and \eqref{eq: MCWF evolution 3}:
\begin{align}
    \hat{\rho}(t+dt)=\mathcal{P}_1 \vert \phi_1 \rangle \langle \phi_1 \vert + \mathcal{P}_2 \vert \phi_2 \rangle \langle \phi_2 \vert+ \mathcal{P}_3 \vert \phi_3 \rangle \langle \phi_3 \vert. \label{eq: density matrix from MCWF}
\end{align}
At time $t$ the density matrix of the system is of the form $\rho(t)=\vert \phi(t) \rangle \langle \phi(t) \vert$. Inserting the results discussed above in \eqref{eq: density matrix from MCWF} and only keeping the terms linear in $dt$ we get
\begin{align}
    \hat{\rho}(t+dt)=\frac{1}{1-\Gamma_e \rho_{ee} dt}\bigg[\hat{\rho}(t) + \mathcal{L}^{\text{tot}}\left\{\hat{\rho}(t)\right\}dt \bigg] \label{eq: density matrix MCWF 2}
\end{align}
with the total linear super-operator acting on the density matrix:
\begin{align}
     \mathcal{L}^{\text{tot}}\left\{\hat{\rho}\right\}=
     &- \frac{i}{\hbar} [\hat{H}_{\text{in}}, \hat{\rho}] 
    + \mathcal{L}^{\text{spon}}\left\{\hat{\rho}\right\}\\ &+\mathcal{L}^{\text{deph}}_5\left\{\hat{\rho}\right\} +\mathcal{L}^{\text{deph}}_e\left\{\hat{\rho}\right\},
\end{align}
in which the first term captures the Hermitian evolution and the non-Hermitian super-operators act as
\begin{align}
    &\mathcal{L}^{\text{spon}}\left\{\hat{\rho}\right\}= 
    \begin{pmatrix}
    &0 &0 &-\frac{\Gamma_e}{2} \rho_{1e} \\
    &0 &0&-\frac{\Gamma_e}{2} \rho_{5e} \\
    &-\frac{\Gamma_e}{2} \rho_{e1} &-\frac{\Gamma_e}{2} \rho_{e5} &-\Gamma_e \rho_{ee}
    \end{pmatrix},
    \\
    &\mathcal{L}^{\text{deph}}_5\left\{\hat{\rho}\right\}= 
    \begin{pmatrix}
    &0 &-\frac{\gamma_5}{2} \rho_{15} &0 \\
    &-\frac{\gamma_5}{2} \rho_{51} &0&-\frac{\gamma_5}{2} \rho_{5e} \\
    &0 &-\frac{\gamma_5}{2} \rho_{e5} &0
    \end{pmatrix},
    \\
    &\mathcal{L}^{\text{deph}}_e\left\{\hat{\rho}\right\}=
    \begin{pmatrix}
    &0 &0 &-\frac{\gamma_e}{2} \rho_{1e} \\
    &0 &0&-\frac{\gamma_e}{2} \rho_{5e} \\
    &-\frac{\gamma_e}{2} \rho_{e1} &-\frac{\gamma_e}{2} \rho_{e5} &0
    \end{pmatrix}.
\end{align}
We note that the super-operator $\mathcal{L}^{\text{spon}}$, corresponding to the spontaneous emission through the excited state is not a trace-preserving operator. Due to the fact that we conditioned the evolution on the absence of  spontaneous emission, the prefactor before the square bracket in \eqref{eq: density matrix MCWF 2} normalizes the loss of population due to $\mathcal{L}^{\text{spon}}$. This factor is reminiscent of the factor \eqref{eq: normalization MCWF tot}. By keeping only the linear  terms in $dt$ {in \eqref{eq: density matrix MCWF 2}}, we obtain the master equation
\begin{equation}
    \boxed{\dv{\hat{\rho}}{t}=\mathcal{L}^{\text{tot}}\left\{\hat{\rho}\right\} + \Gamma_e \rho_{ee} \hat{\rho}}. \label{eq: master equation via MCWF}
\end{equation}
Note that we derived this evolution equation with the assumption that at time $t$ the atom is in a pure internal state. We can use the linearity of the operator $\mathcal{L}^{\text{tot}}$ in \eqref{eq: density matrix MCWF 2} to show the validity of the result \eqref{eq: master equation via MCWF} for the density matrix of a general mixed state.

\subsection{Numerical treatment of the evolution}\label{sec: numerical solutions}
To treat the time-dependent state of the atom, we use the master equation
\begin{equation}
    \dv{\hat{\rho}_{\text{tot}}}{t}=\frac{1}{i\hbar} \left[\hat{H}, \hat{\rho}_{\text{tot}}\right] + \sum_{j}{ \left(\hat{\mathcal{S}}_j\hat{\rho}_{\text{tot}}\hat{\mathcal{S}}_j^\dagger- \frac{1}{2}\left\{\hat{\mathcal{S}}_j^\dagger \hat{\mathcal{S}}_j, \hat{\rho}_{\text{tot}}\right\}\right)}, \label{eq: master equation 6*6}
\end{equation}
in which the total density matrix is of the $6\times6$ form
\begin{align}
    \hat{\rho}_{\text{tot}}= \begin{pmatrix}
    \hat{\rho}_{\text{in}} &\hat{\rho}_{\text{coh}}  \\
    \hat{\rho}_{\text{coh}}^\dagger & \hat{\rho}_{\text{out}}
    \end{pmatrix}, 
\end{align}
where $\hat{\rho}_{\text{in}}$ ($\hat{\rho}_{\text{out}}$) is the $3\times3$ internal state density matrix of the atom that survives (is lost) considering the possibility of spontaneous emission events. Furthermore, the off-diagonal $3\times3$ matrix $\hat{\rho}_{\text{coh}}$ accounts for the quantum coherence between the survived and lost parts of the motional state of the atom. Because the atom is initially purely inside the system ($\hat{\rho}_{\text{out}}=0$) and since the only process that can result in particle loss is spontaneous emission [jumps \eqref{eq: spont e-5 in-out jump} and \eqref{eq: spont e-1 in-out jump}], we can deduce that for all times $\hat{\rho}_{\text{coh}}=0$. Here, we only consider the evolution of $\hat{\rho}_{\text{in}}$, which is independent of $\hat{H}_\Lambda '$. Therefore, this Hamiltonian is irrelevant and we set $\hat{H}_\Lambda '=0$. 
We numerically solve the evolution equation \eqref{eq: master equation 6*6} using the linear MCWF propagation tool provided by the python package \emph{qutip} \cite{johansson_qutip_2012}. In this way, we extract the normalized density matrix of the survived atoms at each time step by
\begin{equation}
    \hat{\rho}(t)=\frac{\hat{\rho}_{\text{in}}(t)}{\Tr{\hat{\rho}_{\text{in}}(t)}}. \label{eq: normalized density matrix}
\end{equation}
Although we discussed the derivation of \eqref{eq: master equation via MCWF} using MCWF approach above, we note that another method of deriving this {master} equation is by inserting the explicit form of the jump operators and the Hamiltonian in \eqref{eq: master equation 6*6} and calculating the time evolution of the normalized density matrix \eqref{eq: normalized density matrix} \cite{dalibard_wave-function_1992, molmer_monte_1993}. 

\subsection{Analytic treatment of the steady state}
The master equation of the $\Lambda$ system in the absence of particle loss is linear \cite{fleischhauer_electromagnetically_2005}. In contrast, we see that the formulation of this loss results in a second order evolution equation \eqref{eq: master equation via MCWF}. Here, we aim for an analytic, steady-state solution for the atoms that remain in the system. In our experiment we have a low probe coupling such that $\Omega_p/\Gamma_e \ll 1$. To solve the problem in this regime, we first write the evolution equation in a dimensionless form by dividing it by $\Gamma_e$. Then we solve it to the lowest order in $\Omega_p/\Gamma_e$ using Wolfram Mathematica. The analytic solutions are presented in section  \ref{section: analytic solution}.

\subsection{Validity of the steady state solution}\label{sec: Lambda system, discussions}
As in our model the loss of an atom can be triggered by any scattering event, we can write the particle loss rate of the atoms initially in $\ket{\downarrow}$ as 
\begin{equation}
    \gamma_{\Lambda}= \Gamma_e \rho_{ee}. \label{eq: Lambda loss rate}
\end{equation}
Since we are most interested in this loss rate (see section \ref{sec: Lambda model}), we assess the convergence time of the density matrix element $\rho_{ee}$ to its steady state. With the convergence criterion $\rho_{ee}(t+dt)-\rho_{ee}(t)<10^{-6}$, we see that the numerical result converges to the analytic steady state solution with a time constant with an upper bound $t_c<5\ \micro\text{s}$ in all experimental parameter regimes discussed in this work. To justify the use of the steady state solution, we compare this time scale with a lower bound of the total time the atom spends in the $\Lambda$ region. In our cold Fermi gas, the velocity of the fastest atoms is on the order of  the Fermi velocity $v_F=\sqrt{2E_F/m}\approx 33.6\ \micro\text{m}/\text{ms}$. On the other hand, the length of the $\Lambda$ region can be estimated using the measured $1/e^2$ waist (see section \ref{sec: ab-initio calibration}): $d\sim2 w_y\approx 3\ \micro\text{m}$. Therefore, a lower bound of the flight time of the atom through the $\Lambda$ region is $\tau>d/v_F\approx 91 \ \micro\text{s}$ showing $\tau \gg t_c$.
Hence, in the majority of the flight time through the beams, the scattering rate obeys the steady state solution. We mention that the local velocity approaching the center of the channel could be much smaller than $v_F$. Therefore, $\tau$ could assume much larger values than $d/v_F$ as we also see from the result of section \ref{sec: Fitting procedure}.

\section{Phenomenological model of transport through channel}\label{sec: Lambda model}
\begin{figure}[t]
    \includegraphics[width=0.5\textwidth]{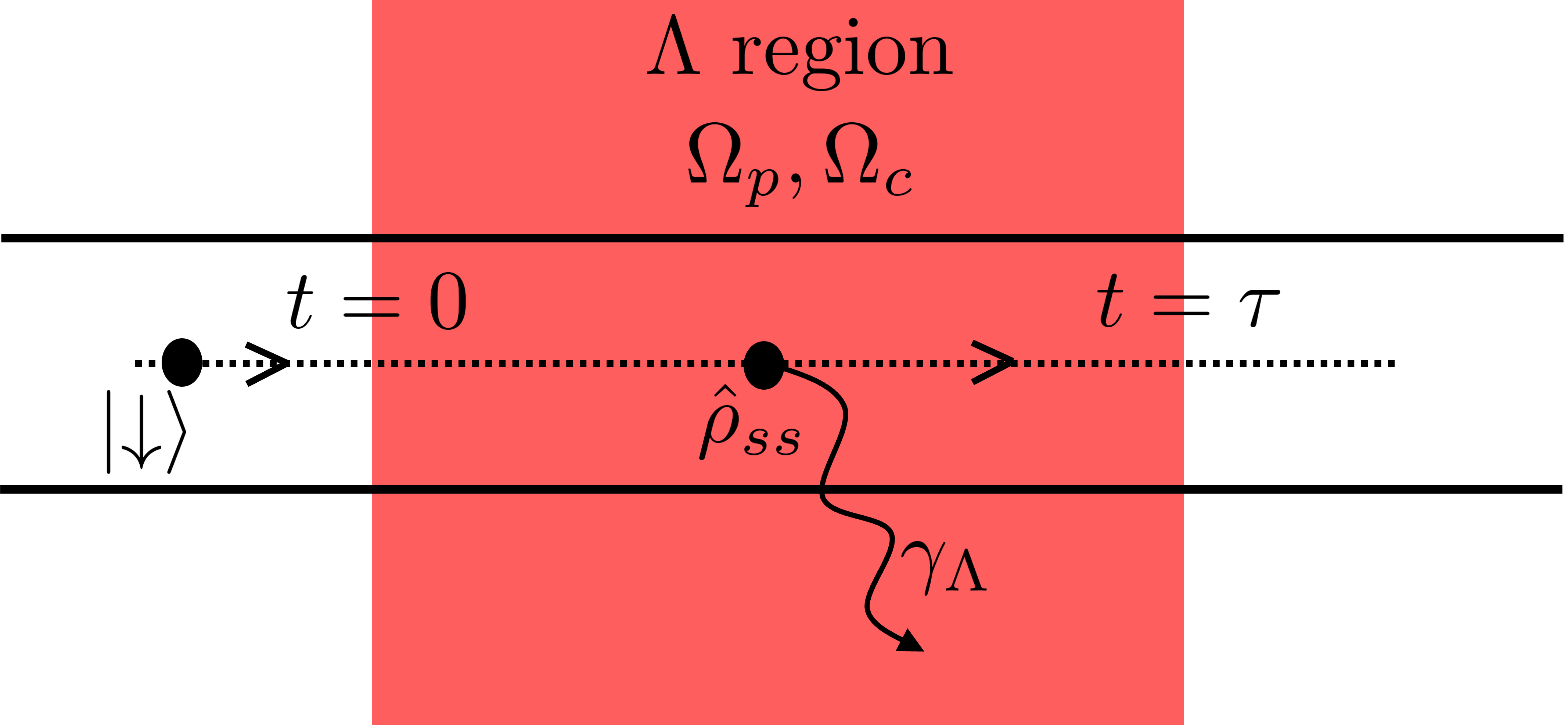}
    \caption{ A schematic of the transport of an atom inside the 1D channel which is initially in state $\kdn$ through the $\Lambda$ region (shown in red). In the $\Lambda$ region, the coupling to common-mode probe and control beams is modelled by effective constant Rabi freqeuencies $\Omega_p$ and $\Omega_c$ respectively. The total flight time is denoted by $\tau$. The particle loss rate of the atom inside the coupling region is modeled by scattering rate $\gamma_\Lambda$ defined in \eqref{eq: Lambda loss rate} for the steady state $\hat{\rho}_{ss}$ of the survived atom calculated from the master equation \eqref{eq: master equation via MCWF}. We adopted this steady state solution as the convergence time $t_c$ is much shorter than the flight time $\tau$ (see section \ref{sec: Lambda system, discussions}).}
    \label{fig: Lambda region schematics}
\end{figure}

In this section, we formulate a model of transport of atoms through the quasi-1D channel incorporating the local $\Lambda$ coupling. The result of this model is used in the next section to fit the measured EIT spectra. In this formulation, the transport of atoms is modeled classically with the average flight time $\tau$ through the $\Lambda$ region. Furthermore, here, we simplify the problem by assuming that each $\kdn$  atom experiences a constant effective Rabi coupling $\Omega_p$ ($\Omega_c$) of the probe (control) beam during the flight. The particle loss rate induced by the $\Lambda$ system during the flight time is modeled by $\gamma_\Lambda$ calculated from the steady state solution of \eqref{eq: master equation via MCWF} (see section \ref{sec: Lambda system, discussions}). A schematic of our model is illustrated in Fig.~\ref{fig: Lambda region schematics}.  The time evolution of the probability of survival in the $\Lambda$ region obeys
\begin{align}
    & \dv{p_\downarrow}{t}=-(\gamma_{\Lambda}+\gamma_{\text{pair}}) p_\downarrow, \label{eq: survival probability 1 ode}\\
    & \dv{p_\uparrow}{t}=-(r_{\text{spin}}\gamma_{\Lambda}+\gamma_{\text{pair}}) p_\downarrow, \label{eq: survival probability 3 ode} \\
    & p_{\downarrow}(0)=p_{\uparrow}(0)=1,
\end{align}
where $p_\downarrow$ ($p_\uparrow$) is the probability of survival of the atom in state $\kdn$ ($\kup$) up to time $t$ during the passage. We assume initial probabilities of one since the atoms enter the $\Lambda$ region as a pure state in either of the spins.
The parameter $\gamma_{\text{pair}}$ is the rate of pair losses, originating in the off-resonant photoassociations induced by the control beam (see section \ref{sec: Photoassociation}). Furthermore, although $\gamma_{\Lambda}$ originates from a $\Lambda$ configuration that does not address atoms in state $\kup$, due to strong interaction between $\kdn$ and $\kup$, each spin-selective loss of $\kdn$ results in a probabilistic loss of atoms in state $\kup$ \cite{huang_superfluid_2023} modeled with the ratio $r_{\text{spin}}$. Due to the small singlet scattering length in ${}^6\text{Li}$~\cite{bartenstein_precise_2005, chin_feshbach_2010}, we neglect the interaction in combinations $\kdn -\vert \text{aux}\rangle$ (1-5) and $\kup -\vert \text{aux}\rangle$ (3-5), similar to the reported 2-6~\cite{van_abeelen_sympathetic_1997} and 1-4~\cite{li_observation_2024} scattering.
By integrating the equations \eqref{eq: survival probability 1 ode} and \eqref{eq: survival probability 3 ode} we get the survival probabilities at the exit time $t=\tau$:
\begin{align}
    &p_\downarrow(\tau)= e^{-(\gamma_{\Lambda}+\gamma_{\text{pair}})\tau}, \label{eq: p_down} \\
    &p_\uparrow(\tau)=1 -  \frac{r_{\text{spin}} \gamma_{\Lambda} +\gamma_{\text{pair}}}{\gamma_{\Lambda}+\gamma_{\text{pair}}}\left(1-e^{-(\gamma_{\Lambda}+\gamma_{\text{pair}})\tau}\right). \label{eq: p_up}
\end{align}
In the absence of $\Lambda$ couplings, we use the notation $N_{\text{tot}}^{(\downarrow/\uparrow)}$ to indicate the total atom number in the cloud for each spin state. Furthermore, we denote the number of the atoms entering the $\Lambda$-region during the beams' total illumination time, which is the same as the transport time $t$, by $N_{\text{pass}}^{(\downarrow/\uparrow)}$ for each state. We assume that {it is a fraction of} the total atom number {given by a spin-independent and} time-dependent phenomenological coefficient 
$f_{\text{pass}}$:
\begin{equation}
    N_{\text{pass}}^{(\downarrow/\uparrow)}= f_{\text{pass}}(t) N_{\text{tot}}^{(\downarrow/\uparrow)}. \label{eq: N_pass assumption}
\end{equation}
Multiplying $N_{\text{pass}}^{(\downarrow/\uparrow)}$ by the loss probability through the channel $1-p_{\downarrow(\uparrow)}(\tau)$ from \eqref{eq: p_down} and \eqref{eq: p_up} results in the accumulated number of lost atoms in each spin state after the total illumination time. By subtracting this lost fraction from the total initial atom number and using \eqref{eq: N_pass assumption}, we get the number of surviving atoms, measured at the end of the experiment: 
\begin{align}
    &N_\downarrow(t)= N_{\text{tot}}^{(\downarrow)}\left[1- f_{\text{pass}}(t) \left(1- p_\downarrow(\tau)\right)\right], \label{eq: N_down}\\
    &N_\uparrow(t)= N_{\text{tot}}^{(\uparrow)}\left[1- f_{\text{pass}}(t) \left(1- p_\uparrow(\tau)\right)\right]. \label{eq: N_up}
\end{align}
By introducing the initial spin ratio $s_0\coloneqq N_{\text{tot}}^{(\downarrow)}/N_{\text{tot}}^{(\uparrow)}$, dividing \eqref{eq: N_down} by \eqref{eq: N_up}, and inserting \eqref{eq: p_down} and \eqref{eq: p_up}, we arrive at the final result for the measured spin ratio $s(t)=N_\downarrow(t)/{N_\uparrow(t)}$:
\begin{align}
    s(t)=
    \frac{s_0\left[1 - f_{\text{pass}}(t) \left( 1-  e^{-(\gamma_{\Lambda}+\gamma_{\text{pair}}) \tau} \right) \right] }{ 1 -  f_{\text{pass}}(t) \frac{r_{\text{spin}} \gamma_{\Lambda} +\gamma_{\text{pair}} }{\gamma_{\Lambda}+\gamma_{\text{pair}}} \left( 1- e^{-(\gamma_{\Lambda}+\gamma_{\text{pair}}) \tau} \right)  }. \label{eq: sratio final formula for fit}
\end{align} 

\section{Fitting procedures}\label{sec: Fitting procedure}
\begin{figure}[t]
    \includegraphics[width=0.5\textwidth]{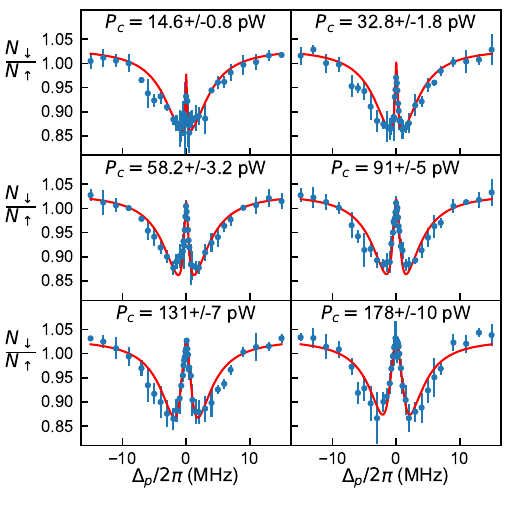}
    \caption{Results of the fits for the observed EIT from the loss spectroscopy of the spin ratio after total illumination time $t=2\ \text{s}$ of the $\Lambda$ beams. The calibrated power of the probe beam is {$P_p=9.1(5) \ \text{pW}$} for all datasets with different control powers $P_c$ (See section \ref{sec: ab-initio calibration} for the calibration of these powers). After applying physical constraints and experimental calibrations (see text), the fits are performed with three free parameters $\Omega_p$, $f_{\text{pass}}(t)$, and $\Delta_c$. }
    \label{fig: fig2 supplement}       
\end{figure}

For the loss spectroscopy of the dark state resonance, we use the equation \eqref{eq: sratio final formula for fit} and fit it to the experimentally measured spin ratio after the illumination time $t=2~\text{s}$. The measurements are performed for six different control powers $P_c$ with a common probe power $P_p=9.1(5)\ \text{pW}$ as illustrated in Fig.~\ref{fig: fig2 supplement}. We fix the ratio $\frac{\Omega_c}{\Omega_p}=\frac{d_c}{d_p} \sqrt{\frac{P_c}{P_p}}$ in all of the data sets with $d_p$ ($d_c$) denoting the dipole element of the probe (control) atomic transition. The ratio of the dipole elements is $d_c/d_p=\sqrt{\Gamma_{e5}/\Gamma_{e1}}=18.51$. Hence, during the fitting, all the control Rabi couplings $\Omega_c$ are constrained to a function of a single free probe Rabi coupling $\Omega_p$. In \eqref{eq: sratio final formula for fit}, there are a few unknowns that are separately determined. As explained
in subsection \ref{sec: calibration of r_spin}, the parameter $r_\text{spin}$ is calibrated from separate measurements. Furthermore, in subsection \ref{sec: calibration of gamma_pair}, $\gamma_{\text{pair}}$ and $\tau$ are separately calibrated as functions of {the fit parameter} $\Omega_p$. We further fix {$s_0=1.03$} to the average {value} at $t=0$ from the transport data used in subsection \ref{sec: calibration of r_spin}. As discussed in section \ref{sec: Conditional Lambda system} the dephasing parameters of the processes \eqref{eq: dephasing 5 jump} and \eqref{eq: dephasing e jump} are fixed to $\gamma_{5}=2\pi\times 0.08\ \text{MHz}$ and $\gamma_e=2\pi\times 1.5\ \text{MHz}$ respectively. Moreover, $\Delta_p$ is calibrated by the observed loss resonance of this beam when the control power is zero. After using all these constraints and calibrations, the fitting is implemented with a minimal number of three free parameters:
\begin{align}
    \ \Omega_p, \ f_{\text{pass}}(t), \ \Delta_c.
\end{align}
We fit all six datasets simultaneously sharing the same fit parameters. The fit residual of each data point (mean value of 4-5 repetitions) is weighted by the inverse of the standard deviation (std) of the repetitions, while we set a  maximum weight using the median std of all data points to improve fit stability. The results give a goodness of fit $\chi^2_{\text{red}}=0.66$ and are shown in Fig.~\ref{fig: fig2 supplement}. From these results we presented three with control powers $P_c\in\{14.6(8),\ 58(3),\ 178(10)\}$~pW in the main text (Fig.~2). The fitted parameters are $\Omega_p=2\pi\times 52(1)\ \text{kHz}$, $f_{\text{pass}}(t)=0.653(5)$, $\Delta_c=2\pi\times 0.03(1)\ \text{MHz}$. We mention that inserting $\Omega_p$ in the calibrated relation \eqref{eq: calibration tau formula} results in the flight time $\tau=330(120)\ \micro \text{s}$. 

\subsection{Calibration of $r_{\text{spin}}$} \label{sec: calibration of r_spin}
\begin{figure}[t]
    \includegraphics[width=0.5\textwidth]{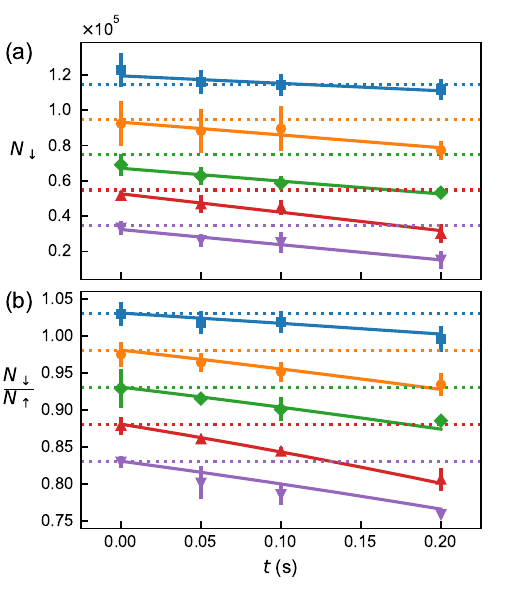}
    \caption{The calibration procedure of the constant $r_{\text{spin}}$. Here, we investigate the short time evolution of the system where only probe beam is present ($P_p>0,\ P_c=0$). In each panel the probe power is set to $P_p=14.6(8), \ 33(2), \ 58(3), 91(5), \ 131(7)\ \text{pW}$, top to bottom respectively. (a) In the first step, equation \eqref{eq: N_down_linearized} is fitted over $N_{\downarrow}$ to extract $N_{\text{tot}}^{(\downarrow)}$ and $\xi$ for each data set. For clarity, the results for different sets are shifted downward vertically. The dashed lines represent the baseline of each curve: the average measured initial atom number $N_{\downarrow}(t=0)=1.2(1)\times 10^5$. (b) In the second step, we fit equation \eqref{eq: sratio_linearized} to all the data sets simultaneously, with a single free parameter $r_\text{spin}$, see text. For clarity, different datasets are shifted downward in the increasing order of $P_p$. For each curve, the dashed lines represent the measured baseline $s_0=1.03(1)$. The final fit result is $r_{\text{spin}}=0.63(2)$. 
    }
    \label{fig: calibrate r_spin}
\end{figure}

To calibrate the parameter $r_{\text{spin}}$, we analyze measurements of the time evolution of $N_{\downarrow}$ and $N_{\uparrow}$ in short time limit when only the power of the probe beam is nonzero ($P_c=0$, $P_p>0$) and it is kept on resonance ($\Delta_p=0$).
Since $P_c=0$, the photoassociation loss is absent ($\gamma_{\text{pair}}=0$) and we can rewrite the survival probabilities \eqref{eq: p_down} and \eqref{eq: p_up} by
\begin{align}
    &p_{\downarrow}=e^{-\gamma_{\text{pump}} \tau}, \label{eq: p_down_PCT2=0}\\
    &p_{\uparrow}=1-r_{\text{spin}}(1-e^{-\gamma_{\text{pump}} \tau}),\label{eq: p_up_PCT2=0}
\end{align}
in which $\gamma_{\Lambda}$ in the absence of the control beam is denoted by $\gamma_{\text{pump}}$. We note that this is the same particle loss mechanism due to optical pumping to state $\vert\text{aux}\rangle$ used in \cite{huang_superfluid_2023} and the orange circles in Fig.~3. Moreover, in the short time limit, we assume that the number of atoms that enter the $\Lambda$ region mentioned in \eqref{eq: N_pass assumption} is linearly dependent on time:
\begin{equation}
    f_{\text{pass}}(t) = f_0 t\, \Rightarrow \ N_{\text{pass}}^{(\downarrow/\uparrow)}= f_0 t\cdot N_{\text{tot}}^{(\downarrow/\uparrow)}. \label{eq: linearize N_pass}
\end{equation}
By inserting equations \eqref{eq: p_down_PCT2=0}, \eqref{eq: p_up_PCT2=0}, and \eqref{eq: linearize N_pass} into \eqref{eq: N_down} and \eqref{eq: N_up} we get the linearized equations for the atom number evolution
\begin{align}
    &N_\downarrow(t)= N_{\text{tot}}^{(\downarrow)}\left(1- \xi t\right), \label{eq: N_down_linearized}\\
    &N_\uparrow(t)= N_{\text{tot}}^{(\uparrow)}\left(1- r_{\text{spin}} \xi t\right), \label{eq: N_up_linearized} 
\end{align}
where we defined
\begin{equation}
    \xi=f_0(1-e^{-\gamma_{\text{pump}} \tau}). \label{eq: definition xi}
\end{equation}
Therefore, we can rewrite the spin ratio in time:
\begin{equation}
    s(t)=\frac{N_\downarrow(t)}{N_\uparrow(t)}=s_0 \frac{1-\xi t}{1 - r_{\text{spin}} \xi t }. \label{eq: sratio_linearized}
\end{equation}
To calibrate $r_{\text{spin}}$, we perform a two-step analysis. In the first step, we fit the equation \eqref{eq: N_down_linearized} to the experimentally measured $N_{\downarrow}$ for five data sets with distinct $P_p$. These fits are implemented with free parameters $N_{\text{(tot)}}^{\downarrow}$ and $\xi$ that are separately extracted for each set [Fig.~\ref{fig: calibrate r_spin}(a)]. 
In the second step, we fit equation \eqref{eq: sratio_linearized} to the spin ratio from the same measurements, with $\xi$ fixed to the fit results from the first step. Furthermore, {$s_0=1.03(1)$} {is fixed to the average} spin ratio at $t=0$. In the end, the fit is performed with a single free parameter $r_{\text{spin}}$ {common to all data sets}. The fit results are presented in Fig.~\ref{fig: calibrate r_spin}(b), with
\begin{equation}
    \boxed{r_{\text{spin}}=0.63(2)},
\end{equation}
which is comparable to the value {$0.72(4)$} calibrated with a different method in a previous work in a similar setting \cite{huang_superfluid_2023}.

\subsection{Calibration of $\gamma_{\text{pair}}$ and $\tau$} \label{sec: calibration of gamma_pair}

The photoassociation pair loss rate $\gamma_{\text{pair}}$ (introduced in section \ref{sec: Lambda model}) is modeled to be only a function of the control power, $\gamma_\text{pair}\propto P_c$. The goal is to calibrate this proportionality factor with experiments only having the control beam, $P_p=0,~P_c>0$, under which condition $\gamma_\Lambda=0$. However, the transport setting introduces a few unknown factors (see below). The idea is then to compare the atom loss rate to that in the complementary condition $P_p>0, P_c=0$, also in the same transport setting. In the latter condition, the scattering rate is determined by the theoretically known optical pumping rate of atoms from $\kdn$ to $\vert\text{aux}\rangle$ denoted by $\gamma_\text{pump}$. Therefore we can relate $\gamma_\text{pair}$ to $\gamma_\text{pump}$, hence to the laser powers. Since we are working in the limit $\Omega_p \ll \Gamma_e$, $\gamma_\text{pump}$ is given by
(see equations \eqref{eq: rho_ee with dephasing} for the case of $\Omega_c=0,\Delta_p=0$ and the discussion of continuous quantum Zeno effect in \cite{streed_continuous_2006}):
\begin{equation}
    \gamma_{\text{pump}}=\frac{\Omega_p^2}{\Gamma_e+\gamma_e}, \label{eq: pumping rate in the continuous Zeno regime}
\end{equation}
where we can see $\gamma_\text{pump} \propto P_p$. Given that $\gamma_{\text{pair}}\propto P_c$, it can be expressed as
\begin{equation}
    \gamma_{\text{pair}}(P_c)=\alpha \frac{P_c}{P_p} \gamma_{\text{pump}}(P_p) , \label{eq: gamma_pair calibration logic}
\end{equation}
where $\alpha$ is the proportionality factor that we aim to experimentally calibrate. This is done in three steps as follows.
\subsubsection*{Step 1}\label{sec: mol step1}
\begin{figure}[t]
    \includegraphics[width=0.5\textwidth]{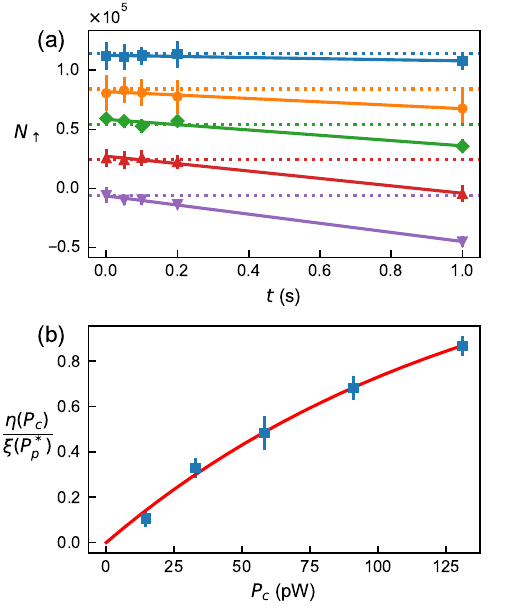}
    \caption{The calibration procedure for the pair loss $\gamma_{\text{pair}}$. (a) Analysis of the atom number for $\kup$ atoms for the power settings $P_p=0$ and $P_c=14.6(8), \ 33(2), \ 58(3), \ 91(5), \ 131(7)\ \text{pW}$ in the increasing order from top to bottom. Except for the smallest $P_c$ (blue square), other datasets are shifted for clarity. The dashed line for each set is the reference atom number fixed to the average $N_\uparrow=1.1(1)\times 10^5$ calculated over all datasets at $t=0$. By fitting the linear equation \eqref{eq: N_up_linearized_mol} over each dataset separately, we extract the parameters $\eta$ and $N_{\text{tot}}^{(\uparrow)}$ for each $P_c$ setting. (b) Ratio $\eta(P_c)/\xi(P^*_p)$ for each {$P_c$, with $\xi(P^*_p)$} extracted from the linear regime ($t\leq1$ s) of the $N_\downarrow$ evolution for the dataset with $P^*_p=14.6(8)\ \text{pW}$ and $P_c=0$ (same data set shown in blue square in Fig.~\ref{fig: calibrate r_spin}). By fitting equation \eqref{eq: l fit function}, we eventually relate $\gamma_\text{pair}$ to beam powers, see text.}
    \label{fig: Fit N_2 with linear mol}
\end{figure}

In the first step, we use the measured atom number evolution with condition $P_p=0$ and $P_c>0$. Under this condition, the probability of survival of atoms in both states from equations \eqref{eq: p_up} and \eqref{eq: p_down} reduces to
\begin{equation}
    p_{\downarrow}=p_{\uparrow}= e^{-\gamma_{\text{pair}} \tau}. 
\end{equation}
Due to the equal loss rate in both spins, here we choose to only analyze the evolution of the  atom number in state $\kup$. With the help of equation \eqref{eq: N_up} and by assuming linear relation \eqref{eq: linearize N_pass} for the short time limit, we arrive at 
\begin{align}
    &N_\uparrow(t)= N_{\text{tot}}^{(\uparrow)}\left(1- \eta t\right), \label{eq: N_up_linearized_mol} 
\end{align}
in which we {defined} 
\begin{equation}
    \eta=f_0(1-e^{-\gamma_{\text{pair}} \tau}). \label{eq: definition eta}
\end{equation}
We extract the parameters $N_{\text{tot}}^{(\uparrow)}$ and $\eta$ from fitting separately to the measurements of $N_{\uparrow}$ with each $P_c$ setting {for $t\leq 1$~s}. Results are presented in Fig.~\ref{fig: Fit N_2 with linear mol}(a), showing that the linearization \eqref{eq: linearize N_pass} is justified in this regime. As $f_0$ and $\tau$ are unknown, it requires additional data to determine $\gamma_\text{pair}$.
\subsubsection*{Step 2}\label{sec: mol step2}
In the second step, we use the data set with the lowest probe power $P_p=14.6(8)\ \text{pW}$ (defined as $P^*_p$) and $P_c=0$. This is the same data set presented as blue squares in Fig.~\ref{fig: calibrate r_spin}. Due to the weak power of the probe beam, we use all the measured time points up to $t=1~\text{s}$ (not all shown in Fig.~\ref{fig: calibrate r_spin}), which fall in the linear evolution regime \eqref{eq: N_down_linearized}. By fitting equation \eqref{eq: N_down_linearized} as discussed in section \ref{sec: calibration of r_spin}, we extract the parameter $\xi(P^*_p)=0.39(2)$. We also define the corresponding loss rate as $\gamma^*_\text{pump}$. Therefore, from \eqref{eq: gamma_pair calibration logic}, $\gamma_\mathrm{pair}$ can be written as 
\begin{equation}
    \gamma_\mathrm{pair}=\alpha \gamma^*_\text{pump} P_c/P^*_p 
    \label{eq: gamma_pair calibration 2}
\end{equation}

\subsubsection*{Step 3}
Within our model, we assume that $f_0$ and $\tau$ in $\xi$ of \eqref{eq: definition xi} are the same as those in $\eta$ of \eqref{eq: definition eta}. Therefore, from the previously fitted $\eta(P_c)$ (step 1) and $\xi(P^*_p)$ (step 2), we can eliminate the unknown factor $f_0$ and obtain
\begin{equation}
    \frac{\eta(P_c)}{\xi(P^*_p)}=\frac{1-e^{-\gamma_{\text{pair} } \tau}}{1- e^{-\gamma^*_{\text{pump}} \tau}}=\frac{1-e^{-A^*\cdot \alpha P_c/P^*_p}}{1- e^{-A^*}}, \label{eq: l fit function}
\end{equation}
where we defined $A^*\coloneqq \gamma_{\text{pump}}^* \tau$ and used \eqref{eq: gamma_pair calibration 2}.
We use equation \eqref{eq: l fit function}
and fit it to the obtained $\eta(P_c)$ from step 1 as a function of $P_c$ [see Fig.~\ref{fig: Fit N_2 with linear mol}
(b)], which results in
\begin{equation}
\boxed{\alpha=0.086(8)}, \boxed{A^*=1.2(4)}.  \label{eq: alpha and Astar values}
\end{equation}
We can rewrite the equation \eqref{eq: gamma_pair calibration logic} by using \eqref{eq: pumping rate in the continuous Zeno regime} for general power settings:
\begin{align}
    \gamma_{\text{pair}}(P_c)=\alpha \frac{P_c}{P_p} \frac{\Omega_p^2}{\Gamma_e+\gamma_e}, \label{eq: gamma pair calibration final}
\end{align}
in which $\Omega_p$ is the probe Rabi frequency corresponding to the chosen power $P_p$. In this way, we calibrate $\gamma_{\text{pair}}$ in terms of the fit parameter of the full model, $\Omega_p$. We can also extract the flight time $\tau$ from the fitted $A^*$. Again by using \eqref{eq: pumping rate in the continuous Zeno regime}, we can rewrite it in terms of $\Omega_p$ with
\begin{equation}
  \tau=\frac{A^*P_p/P^*_p}{\Omega_p^2/(\Gamma_e+\gamma_e)}. \label{eq: calibration tau formula} 
\end{equation}
We did check that leaving $\tau$ as a free parameter in the final fit using \eqref{eq: sratio final formula for fit} does not much influence the fit quality and has almost no influence on the obtained $\Omega_p$ and $\Omega_c$. We therefore keep the fit as simple as possible by constraining $\tau$ to $\Omega_p$ using  \eqref{eq: calibration tau formula}.

\section{Estimating dipole potentials}
\begin{figure}[t]
    \includegraphics[width=0.5\textwidth]{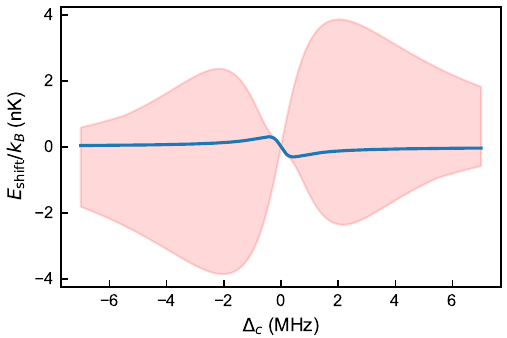}
    \caption{The light shift \eqref{eq: light shift formula} in terms of $\Delta_c$ in the configuration of $\Lambda$-system parameters of the experiment of Fig.~4, red triangles. The solid blue line is the value of the light shift of the steady state solution of \eqref{eq: master equation via MCWF}. The transient light shift before reaching the steady state explores the shaded red region.}
    \label{fig: lightshift bounds}       
\end{figure}
Here we formulate the total dipole potential that an atom feels under the $\Lambda$ coupling.
In a general time-dependent rotating frame 
\begin{equation}
    \vert \tilde{\psi} \rangle = \hat{U}(t) \vert \psi \rangle, \label{eq: state rotation}
\end{equation}
where we denote the parameters in the rotating frame with $\tilde{}$ and the ones in the Schrödinger frame without, the Hamiltonian transforms as
\begin{equation}
    \hat{\tilde{H}}= \hat{U} \hat{H} \hat{U}^\dagger + i\hbar \dv{\hat{U}}{t} \hat{U}^\dagger. \label{eq: Hamiltonian rotation}
\end{equation}
The energy shift of the atom in the state $\vert \psi \rangle$ expressed in the Schrödinger frame is
\begin{equation}
    E_{\text{shift}}=\langle \psi \vert \hat{H} \vert \psi \rangle - \langle \psi \vert \hat{H}_0 \vert \psi \rangle, \label{eq: Energy_shift_in_Schrodinger_frame}
\end{equation}
in which $H_0$ is the Hamiltonian of the atom in the absence of any light field. We can rewrite equation \eqref{eq: Energy_shift_in_Schrodinger_frame} using \eqref{eq: state rotation} and \eqref{eq: Hamiltonian rotation}:
\begin{align}
    E_{\text{shift}} &= \langle \tilde{\psi} \vert \hat{U} (\hat{H}-\hat{H}_0) \hat{U}^\dagger \vert \tilde{\psi} \rangle\\
    &=\langle \tilde{\psi} \vert  \underbrace{\left(\hat{\tilde{H}} -i\hbar \dv{\hat{U}}{t} \hat{U}^\dagger - \hat{U} \hat{H}_0 \hat{U}^\dagger \right)}_{\hat{H}^\prime}  \vert \tilde{\psi}\rangle
.\end{align}
Using the general density matrix $\hat{\rho}=\sum_{j=1}^3 p_j \vert \tilde{\phi}_j \rangle \langle \tilde{\phi}_j \vert$ expressed in the rotating frame, we get the expression for the light shift:
\begin{equation}
    E_{\text{shift}}=\Tr{\hat{H}^\prime \hat{\rho}}. \label{eq: lightshift}
\end{equation}
For the case of the three-level system in the rotating frame with Hamiltonian \eqref{eq: Hamiltonian in} and its corresponding time-dependent unitary operator we have
\begin{equation}
    \hat{H}^\prime = \hbar
    \begin{pmatrix}
    0 &0 &-\frac{\Omega_p}{2} \\
    0 &0 &-\frac{\Omega_c}{2} \\
    -\frac{\Omega_p}{2} &-\frac{\Omega_c}{2} &0 
    \end{pmatrix}.\label{eq: effective Hamiltonian, lightshift}
\end{equation}
We see that this effective Hamiltonian used for the calculation of the light shift is independent of the detuning. However, the detuning could influence the state of the system $\hat{\rho}$ which in turn affects $E_{\text{shift}}$. Inserting \eqref{eq: effective Hamiltonian, lightshift} in \eqref{eq: lightshift} results in
\begin{equation}
    E_\text{shift}=-\hbar \Omega_p \text{Re}\left\{\rho_{1e}\right\}-\hbar \Omega_c \text{Re}\left\{\rho_{5e}\right\}. \label{eq: light shift formula}
\end{equation}
Using the steady state solutions of \eqref{eq: master equation via MCWF} from section \ref{section: analytic solution}, we see that at the single-photon resonance ($\Delta_p=0$, same condition as in Fig.~4), and in the limit of zero two-photon dephasing ($\gamma_5=0$), we have $E_{\text{shift}}=0$ for all two-photon detunings $\delta$ close to the resonance. As $\gamma_5$ increases, $E_{\text{shift}}$ deviates more from zero. To find a maximum bound for the magnitude of $E_{\text{shift}}$, we theoretically analyze this parameter for the case of the maximum bound of the two-photon dephasing in our experiment, $\gamma_5=2\pi\times 80~\text{kHz}$. The results as a function of $\Delta_c$ for $\Lambda$ parameters $\Omega_p=2\pi\times0.098~\text{MHz}$, $\Omega_c=2\pi\times2.42~\text{MHz}$, and $\Delta_p=0$ (same setting as red triangles in Fig.~4)  are presented in Fig.~\ref{fig: lightshift bounds}. Blue solid line shows the analytic steady state solution calculated by inserting \eqref{eq: steady Real_rhoe1} and \eqref{eq: steady Real_rhoe5} in \eqref{eq: light shift formula}. The shaded region shows maximum excursions of all transient values of $E_{\text{shift}}$ the atom experiences before reaching the steady state. For this, the time-dependent value of the density matrix extracted by the numerical method discussed in section \ref{sec: numerical solutions} is inserted in \eqref{eq: light shift formula}. We see that the value of the light shift does not exceed $k_B\cdot4~\text{nK}$ at all times. In comparison with other energy scales of the system such as temperature $T=80(1)~\text{nK}$, Fermi energy $E_F=k_B\cdot408(5)~\text{nK}$, transverse confinement energies of the channel (see section \ref{sec: transport geometry}) $\nu_x^{(\text{ch})}=k_B/h\cdot 592(11)~\text{nK}$, $\nu_z^{(\text{ch})}=k_B/h\cdot 514(6)~\text{nK}$, $E_\text{shift}$ is negligible, suggesting that the observed asymmetry in Fig.~4(a) does not originate from the light shift of the $\Lambda$ system. 

\section{Experimental details}\label{sec: experimental details}

\subsection{Calibration of Rabi frequencies by power measurements}\label{sec: ab-initio calibration}

To extract the intensity of the beams on the atoms, we image the beam profile and fit it to a 2D Gaussian function,
\begin{equation}
    I(x,y)= \frac{2P_0}{\pi w_{x} w_{y}} e^{-\frac{2x^2}{w_{x}^2}} e^{-\frac{2y^2}{w_{y}^2}}, \label{eq: Gaussian_beam}
\end{equation}
where $P_0$ is the beam power, resulting in the waists $w_x=1.38~\micro$m, $w_y=1.52\ \micro$m. The Rabi frequency is related to the intensity via 
\begin{align}
    \Omega/2\pi=\sqrt{\frac{2}{c\epsilon_0h^2}} d_{ge}\sqrt{I},
\end{align}
where $c$ is the speed of light, $\epsilon_0$ is the vacuum permittivity, and $d_{ge}$ is the dipole moment, taking the values \cite{gehm2003preparation} 
\begin{align}
    &d_{p}={1.073\times10^{-30} \ \text{Cm}},\label{eq: dipole probe} \\
    &d_{c}={1.985\times 10^{-29} \ \text{Cm}},\label{eq: dipole control}
\end{align}
for the probe and control transitions respectively.
With this, we calibrate the peak Rabi frequencies for each beam:
\begin{align}
    \Omega_p^{\text{peak}}/2\pi={ 24.43(3) \ \text{kHz}\cdot \sqrt{P_p~[\text{pW}]}}, \label{eq: Rabi_calibration_probe} \\
    \Omega_c^{\text{peak}}/2\pi = {452.2(5) \ \text{kHz}\cdot \sqrt{P_c~[\text{pW}]}} \label{eq: Rabi_calibration_control},
\end{align}
with $P_p$ and $P_c$ the  power of the beams on the atomic cloud in the $\sigma_+$ polarization expressed in $\text{pW}$. Here, we compare the calibrations \eqref{eq: Rabi_calibration_probe} and \eqref{eq: Rabi_calibration_control} with the Rabi couplings resulted from the fit in Fig.~2. There, we experimentally fixed $P_p=9.1(5)~\text{pW}$ which corresponds to $\Omega_p^{\text{peak}}=2\pi\times74(2)~\text{kHz}$ from \eqref{eq: Rabi_calibration_probe}. In the theoretical model that we used for the fit (see section \ref{sec: Lambda model}), the effective Rabi couplings are  constant across the $\Lambda$ region. Therefore, to compare with theory, we extract a constant Rabi coupling $\bar{\Omega}_p$ by averaging the Gaussian profile of the measured intensity in a region with length $2w_y$ along $y$ direction. With this, we get $\bar{\Omega}_p=2\pi\times58(2)~\text{kHz}$ which is in good agreement with $\Omega_p=2\pi\times 52(1)~\text{kHz}$ from the fit result of Fig.~2.

\subsection{Transport geometry and the experimental sequence}\label{sec: transport geometry}

To create the potential landscape for the transport configuration, a cigar-shaped gas trapped in a hybrid magnetic and dipole trap with trapping frequencies $\nu_x=168(2) \ \text{Hz}, \ \nu_y=28.18(4) \ \text{Hz}, \ \nu_z=148(1) \ \text{Hz}$ is partitioned in two reservoirs using two $\text{TEM}_{01}$-like repulsive beams which propagate along $x$ and $z$ directions as defined in Fig.~1. This results in a transport channel along the $y$ direction between the two reservoirs with peak confinement frequencies $\nu_{x}^{(\text{ch})}=12.3(2) \ \text{kHz}, \ \nu_{z}^{(\text{ch})}=10.7(1)\ \text{kHz}$. Since these confinement energies are much larger than the reservoirs' average temperature $k_BT/h=1.67(3)\ \text{kHz}$, the transport happens in the 1D regime. The length $l= 6.9\ \micro \text{m}$ of the 1D region is determined by the smallest waist along $y$ of the TEM$_{01}$-like beams. To tune the number of occupied transverse modes and the degeneracy in the channel, an attractive $767\ \text{nm}$ \emph{gate} beam covering the channel region is used. To fully block the transport through the channel, we use another repulsive \emph{wall} beam propagating along $z$ with an elliptical profile  with a smaller waist along $y$ centered at the channel. Starting from a particle number imbalance between the two reservoirs with the wall beam blocking the transport, we switch off the wall for the time duration $t$ after which it is switched back on. After ramping down all confinement beams to zero, we measure the density profile in each reservoir. In this way, we access the state of the reservoirs after  transport time $t$. See \cite{fabritius_irreversible_2024} for more details on the experimental sequence and the transport geometry of this experiment.
\begin{figure}[t]
    \includegraphics[width=0.5\textwidth]{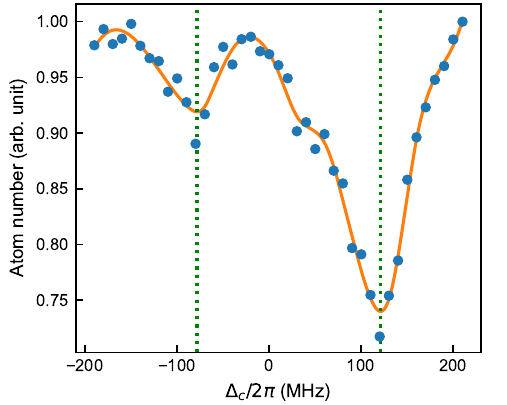}
    \caption{ Normalized photoassociation pair loss induced by the control beam alone. We observe {mainly} two resonances (shown in green dotted lines) at $\Delta_c\sim 121\ \text{MHz}$ and $\Delta_c\sim -78\ \text{MHz}$. The orange solid line is an interpolation of the data points as a guide to the eye.}
    \label{fig: photoassociation loss spectroscopy}
\end{figure}

\subsection{Photoassociation pair loss} \label{sec: Photoassociation}
In the unitary regime, we observe photoassociation pair losses \cite{jones_ultracold_2006, partridge_molecular_2005} induced by the control beam. The measured atom loss spectrum of only the control beam ($P_p=0,~P_c>0$) is presented in Fig.~\ref{fig: photoassociation loss spectroscopy}. We observe mainly two broad resonances detuned from the $\vert\text{aux}\rangle-\vert e \rangle$ resonance ($\Delta_c=0$). These measurements are performed with the control beam located in the reservoir to improve the signal to noise ratio.

\section{Discussions on the asymmetric transport versus two-photon detuning} \label{sec: asymmetry}

\subsection{Measurements at higher temperatures}
\begin{figure}[t]
    \includegraphics[width=0.5\textwidth]{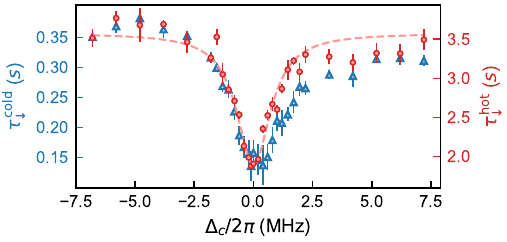}
    \caption{Transport timescales versus control beam detuning at different temperatures of the gas. Data for the cold gas ($\tau_\downarrow^\mathrm{cold}$, blue triangles, left axis) correspond to $T/T_F\approx 0.23$ in the reservoirs, and data for the hot gas ($\tau_\downarrow^\mathrm{hot}$, red circles, right axis) correspond to $T/T_F\approx 0.77$ in the reservoirs. Dashed red curve is a Lorentzian fit to the high temperature data as a guide to the eye. Error bars represent standard errors from the fits. }
    \label{fig: asymmetry_high_temp}
\end{figure}

We have performed similar transport measurements as in Fig.~4 at unitarity while preparing the reservoirs at higher temperatures. The transport timescales $\tau_{\downarrow}$ of $\ket{\downarrow}$ atoms are shown in Fig.~\ref{fig: asymmetry_high_temp}, where $\tau_{\downarrow}$ is the time constant of an exponential fit to $\Delta N_\downarrow/N_{\downarrow}$. We plot $\tau_{\downarrow}$ instead of the average of the two spins shown in Fig.~4 because at high temperatures $\ket{\uparrow}$ atoms become weakly affected by the probe beam and the transport dynamics of the two spins become more distinct. These measurements are taken at Rabi frequencies  $\Omega_p=2\pi\times 0.173(6)$~MHz, $\Omega_c=2\pi\times 3.9(2)$~MHz, which are chosen slightly higher than those in Fig.~4 in order to have a clear signal in the hotter gas.
Compared to the cold gas (blue triangles, left axis), the data at higher temperature (red circles, right axis) show much slower transport and the asymmetry with respect to the control beam detuning $\Delta_c$ diminishes. In addition to the comparison with the non-interacting results in Fig.~4, this observation suggests that the asymmetry is related to pairing between $\ket{\downarrow}$ and $\ket{\uparrow}$.

\subsection{Off-resonant dipole potentials}
Since the probe beam is always on resonance $\Delta_p=0$, and only $\Delta_c$ is varied, we only consider off-resonant transitions for the control laser. This laser is red detuned with respect to the most relevant off-resonant transitions for both spin states, $\ket{\downarrow}\equiv\ket{S_{1/2},-1/2,1}\rightarrow\ket{P_{3/2},1/2,1}$ and $\ket{\uparrow}\equiv\ket{S_{1/2},-1/2,-1}\rightarrow\ket{P_{3/2},1/2,-1}$, where the notations are in $\ket{L_{J},m_J,m_I}$ basis. With increasing $\Delta_c$, which means increasing the red-detuning from these transitions, the attractive potential, if significant, would decrease and slow down the transport. This is the opposite of the observed asymmetry---the transport is faster on the $\Delta_c>0$ side---so cannot explain the asymmetry. 

We can further consider the off-resonant molecular transition shown in Fig.~\ref{fig: photoassociation loss spectroscopy}. As the pair loss seems to be dominated by the resonance features at $\Delta_c>0$, we primarily consider these resonances. Note that the laser is in fact blue-detuned with respect to these resonances given our definition (Fig.~1 inset). Increasing $\Delta_c$ means the laser approaches the resonance from the blue side, increasing the repulsive potential, which should increase the transport timescale on the $\Delta_c>0$ side. This is also the opposite of the observation.  

\onecolumngrid

\section{Analytic form of the steady-state of the density matrix of the surviving atoms in the limit $\Omega_p\ll\Gamma_e$ } \label{section: analytic solution}

Here we present the analytic steady state solutions of the equation \eqref{eq: master equation via MCWF} in the limit $\Omega_p\ll \Gamma_e$. In the density matrix elements, the subscripts $1$ and $5$ correspond to the states $\kdn$ and $\vert \text{aux}\rangle$ respectively. The rest of the notations are all introduced in section \ref{sec: Conditional Lambda system}.

\subsection{Solutions in the general case}
We first define the frequently appearing denominator notations
\begin{align}
    \bm{D_a}=&\left(\gamma_5^2+4 (\Delta_p-\Delta_c)^2\right) \left((\gamma_e+\Gamma_e)^2+4 \Delta_p^2\right)+2 \Omega_c^2 (\gamma_5 (\gamma_e+\Gamma_e)+4 \Delta_p (\Delta_c-\Delta_p))+\Omega_c^4,
    \\
    \bm{D_b}=&(\gamma_5+\gamma_e+\Gamma_e) \bm{D_a}.
\end{align}
The resulting density matrix elements follow
\begin{align}
        \rho_{11}=& 1-\bm{\frac{N_{11}}{D_{b}}}  \left(\frac{\Omega_p}{\Gamma_e}\right)^2 + \text{O}\left[\left(\frac{\Omega_p}{\Gamma_e}\right)^4\right],
\end{align}
in which
\begin{align} 
    \bm{N_{11}}=& \Gamma_e \bigg[ \gamma_5^2 \left((\gamma_e+\Gamma_e) (2 \gamma_e+3 \Gamma_e)+2 \Omega_c^2\right)+2 \gamma_5^3 (\gamma_e+\Gamma_e)+(\gamma_e+\Gamma_e) \left(4 (2 \gamma_e+\Gamma_e) (\Delta_p-\Delta_c)^2+\Gamma_e \Omega_c^2\right) \notag
    \\
    & + \gamma_5 \left(\gamma_e^2 \Gamma_e+2 \gamma_e \left(\Gamma_e^2+4 (\Delta_p-\Delta_c)^2+\Omega_c^2\right)+\Gamma_e \left(\Gamma_e^2+4 \Delta_p^2-8 \Delta_p \Delta_c+8 \Delta_c^2+3 \Omega_c^2\right)\right) \bigg]. 
\end{align}
\begin{align}
    \rho_{55}=&\bm{\frac{N_{55}}{D_{b}}} \left(\frac{\Omega_p}{\Gamma_e}\right)^2 + \text{O}\left[\left(\frac{\Omega_p}{\Gamma_e}\right)^4\right],
\end{align}
in which
\begin{align}
    \bm{N_{55}}=& \Gamma_e \bigg[ \gamma_5^2 \left((\gamma_e+\Gamma_e) (\gamma_e+2 \Gamma_e)+\Omega_c^2\right)+\gamma_5^3 (\gamma_e+\Gamma_e)+(\gamma_e+\Gamma_e) \left(4 \gamma_e (\Delta_p-\Delta_c)^2+\Gamma_e \Omega_c^2\right) \notag
    \\
    &+ \gamma_5 \left(\gamma_e^2 \Gamma_e+\gamma_e \left(2 \Gamma_e^2+4 (\Delta_p-\Delta_c)^2+\Omega_c^2\right)+\Gamma_e \left(\Gamma_e^2+4 \Delta_c^2+2 \Omega_c^2\right)\right) \bigg].
\end{align}

\begin{align}
    \rho_{ee}=&\bm{\frac{N_{ee}}{D_a}} \left(\frac{\Omega_p}{\Gamma_e}\right)^2 + \text{O}\left[\left(\frac{\Omega_p}{\Gamma_e}\right)^4\right], \label{eq: rho_ee with dephasing}
\end{align}
in which
\begin{align}
    \bm{N_{ee}}=& \Gamma_e \left((\gamma_e+\Gamma_e) \left(\gamma_5^2+4 (\Delta_p-\Delta_c)^2\right)+\gamma_5 \Omega_c^2\right). \label{eq: rho_ee num with dephasing}
\end{align}

\begin{align}
    \text{Im}\{\rho_{e1}\}=&\bm{\frac{N_{e1}^{\text{Im}}}{D_a}} \left(\frac{\Omega_p}{\Gamma_e}\right) + \text{O}\left[\left(\frac{\Omega_p}{\Gamma_e}\right)^3\right],
\end{align}
in which
\begin{align}
    \bm{N_{e1}^{\text{Im}}}=& \Gamma_e \left((\gamma_e+\Gamma_e) \left(\gamma_5^2+4 (\Delta_p-\Delta_c)^2\right)+\gamma_5 \Omega_c^2\right).
\end{align}

\begin{align}
     \text{Re}\{\rho_{e1}\}=&\bm{\frac{N_{e1}^{\text{Re}}}{D_a}} \left(\frac{\Omega_p}{\Gamma_e}\right) + \text{O}\left[\left(\frac{\Omega_p}{\Gamma_e}\right)^3\right], \label{eq: steady Real_rhoe1}
\end{align}
in which
\begin{align}
    \bm{N_{e1}^{\text{Re}}}= & 2 \Gamma_e \left(\Delta_p \left(\gamma_5^2+4 (\Delta_p-\Delta_c)^2\right)+\Omega_c^2 (\Delta_c-\Delta_p)\right).
\end{align}

\begin{align}
    \text{Im}\{\rho_{e5}\}=&0+  \text{O}\left[\left(\frac{\Omega_p}{\Gamma_e}\right)^4\right].
\end{align}

\begin{align}
    \text{Re}\{\rho_{e5}\}=\bm{\frac{N_{e5}^{\text{Re}}}{D_{b}}} \left(\frac{\Omega_p}{\Gamma_e}\right)^2 + \text{O}\left[\left(\frac{\Omega_p}{\Gamma_e}\right)^4\right], \label{eq: steady Real_rhoe5}
\end{align}
in which
\begin{align}
    & \bm{N_{e5}^{\text{Re}}} =  2 \Gamma_e^2 \Omega_c (\gamma_5 \Delta_p+(\gamma_e+\Gamma_e) (\Delta_p-\Delta_c)).
\end{align}

\begin{align}
    \text{Im}\{\rho_{51}\}=\bm{\frac{N_{51}^{\text{Im}}}{D_a}} \left(\frac{\Omega_p}{\Gamma_e}\right) + \text{O}\left[\left(\frac{\Omega_p}{\Gamma_e}\right)^3\right],
\end{align}
in which
\begin{align}
    & \bm{N_{51}^{\text{Im}}}= 2 \Gamma_e \Omega_c (\gamma_5 \Delta_p+(\gamma_e+\Gamma_e) (\Delta_p-\Delta_c)).
\end{align}

\begin{align}
    \text{Re}\{\rho_{51}\}=\bm{\frac{N_{51}^{\text{Re}}}{D_a}} \left(\frac{\Omega_p}{\Gamma_e}\right) + \text{O}\left[\left(\frac{\Omega_p}{\Gamma_e}\right)^3\right],
\end{align}
in which
\begin{align}
    & \bm{N_{51}^{\text{Re}}}= -\Gamma_e \Omega_c \left(\gamma_5 (\gamma_e+\Gamma_e)+4 \Delta_p (\Delta_c-\Delta_p)+\Omega_c^2\right).
\end{align}

\subsection{Solutions in the case of zero dephasing}
Here we simplify the results of the previous subsection for the case of zero dephasing $\gamma_5=\gamma_e=0$. We first define the frequently recurring denominator notation
\begin{align}
    \bm{D}=4 \Gamma_e^2 (\Delta_p-\Delta_c)^2+\left(4 \Delta_p (\Delta_c-\Delta_p)+\Omega_c^2\right)^2.
\end{align}
The resulting density matrix elements follow:
\begin{align}
    \rho_{11}=1- \frac{\Gamma_e^2 \left(4 (\Delta_p-\Delta_c)^2+\Omega_c^2\right)}{\bm{D}} \left(\frac{\Omega_p}{\Gamma_e}\right)^2 + \text{O}\left[\left(\frac{\Omega_p}{\Gamma_e}\right)^4\right],
\end{align}

\begin{align}
    \rho_{55}=&\frac{\Gamma_e^2 \Omega_c^2}{\bm{D}}  \left(\frac{\Omega_p}{\Gamma_e}\right)^2 + \text{O}\left[\left(\frac{\Omega_p}{\Gamma_e}\right)^4\right],
\end{align}

\begin{align}
    \rho_{ee}=& \frac{4 \Gamma_e^2 (\Delta_p-\Delta_c)^2}{\bm{D}} \left(\frac{\Omega_p}{\Gamma_e}\right)^2 + \text{O}\left[\left(\frac{\Omega_p}{\Gamma_e}\right)^4\right],
\end{align}

\begin{align}
    \text{Im}\{\rho_{e1}\}=&\frac{4 \Gamma_e^2 (\Delta_p-\Delta_c)^2}{\bm{D}} \left(\frac{\Omega_p}{\Gamma_e}\right) + \text{O}\left[\left(\frac{\Omega_p}{\Gamma_e}\right)^3\right],
\end{align}

\begin{align}
    \text{Re}\{\rho_{e1}\}=& \frac{2 \Gamma_e (\Delta_p-\Delta_c) \left(4 \Delta_p (\Delta_p-\Delta_c)-\Omega_c^2\right)}{\bm{D}} \left(\frac{\Omega_p}{\Gamma_e}\right) + \text{O}\left[\left(\frac{\Omega_p}{\Gamma_e}\right)^3\right],
\end{align}

\begin{align}
     \text{Im}\{\rho_{e5}\}=&0+  \text{O}\left[\left(\frac{\Omega_p}{\Gamma_e}\right)^4\right],
\end{align}

\begin{align}
    \text{Re}\{\rho_{e5}\}=\frac{2 \Gamma_e^2 \Omega_c (\Delta_p-\Delta_c)}{\bm{D}} \left(\frac{\Omega_p}{\Gamma_e}\right)^2 + \text{O}\left[\left(\frac{\Omega_p}{\Gamma_e}\right)^4\right],
\end{align}

\begin{align}
    \text{Im}\{\rho_{51}\}=\frac{2 \Gamma_e^2 \Omega_c (\Delta_p-\Delta_c)}{\bm{D}} \left(\frac{\Omega_p}{\Gamma_e}\right) + \text{O}\left[\left(\frac{\Omega_p}{\Gamma_e}\right)^3\right],
\end{align}

\begin{align}
    \text{Re}\{\rho_{51}\}=-\frac{\Gamma_e \Omega_c \left(4 \Delta_p (\Delta_c-\Delta_p)+\Omega_c^2\right)}{\bm{D}} \left(\frac{\Omega_p}{\Gamma_e}\right) + \text{O}\left[\left(\frac{\Omega_p}{\Gamma_e}\right)^3\right].
\end{align}

\end{document}